%% file: main.tex
\documentclass[fleqn,usenatbib]{mnras}

\usepackage{newtxtext,newtxmath}

\usepackage[T1]{fontenc}
\usepackage[utf8]{inputenc}
\usepackage{ae,aecompl}
\usepackage{enumitem}

\usepackage{graphicx}	
\usepackage[singlelinecheck=off]{subcaption}
\usepackage{amsmath}	
\usepackage{amssymb}	
\usepackage{adjustbox}
\usepackage{nccmath}
\usepackage{color}
\usepackage[displaymath]{lineno}
\usepackage{eso-pic}

\DeclareMathSizes{12}{12}{10}{8}
\title[DES17X1boj and DES16E2bjy]{The Mystery of Photometric Twins DES17X1boj and DES16E2bjy}

\input DES-2019-0495_author_list_only_final2.tex

\date{Accepted XXX. Received YYY; in original form ZZZ}

\pubyear{2019}

\begin{document}
\label{firstpage}
\pagerange{\pageref{firstpage}--\pageref{lastpage}}
\maketitle

\begin{abstract}
We present an analysis of DES17X1boj and DES16E2bjy, two peculiar transients discovered by the Dark Energy Survey (DES).  They exhibit nearly identical double-peaked light curves which reach very different maximum luminosities ($M_\mathrm{r}=-15.4$ and $M_\mathrm{r}=-17.9$, respectively). The light curve evolution of these events is highly atypical and has not been reported before. The transients are found in different host environments: DES17X1boj was found near the nucleus of a spiral galaxy, while DES16E2bjy is located in the outskirts of a passive red galaxy. Early photometric data is well fitted with a blackbody and the resulting moderate photospheric expansion velocities ($1800$ km/s for DES17X1boj and $4800$ km/s for DES16E2bjy) suggest an explosive or eruptive origin. Additionally, a feature identified as high-velocity CaII absorption ($v\approx9400$km/s) in the near-peak spectrum of DES17X1boj may imply that it is a supernova. While similar light curve evolution suggests a similar physical origin for these two transients, we are not able to identify or characterise the progenitors.
\end{abstract}




\section{Introduction}

Dedicated wide-field supernova (SN) surveys have discovered large numbers of extragalactic transients over the last decade. A large majority of them are traditional types of SNe, including type Ia SNe produced in thermonuclear disruptions of white dwarfs and type II/Ibc originating in the aftermath of the core-collapse of massive stars ($\gtrsim8M_{\odot}$, see e.g. \citealt{Filippenko1997}, \citealt{Gal-Yam2017} for reviews). However, due to the increase in area, cadence and depth, these surveys have also started to discover rarer types of events such as the rapidly evolving transients \citep[e.g. ][]{Drout2014, Pursiainen2018}, superluminous supernovae (SLSNe; e.g. \citealt{Quimby2011}; see \citealt{Howell2017} for a review) and Ca-rich transients \citep[e.g. ][]{Perets2010}, all of whose behaviour cannot be explained with the physical mechanisms used for typical SNe. 

\begin{figure*}
    \centering
    \begin{subfigure}[b]{0.45\textwidth}
        \centering
        \includegraphics[width=\textwidth]{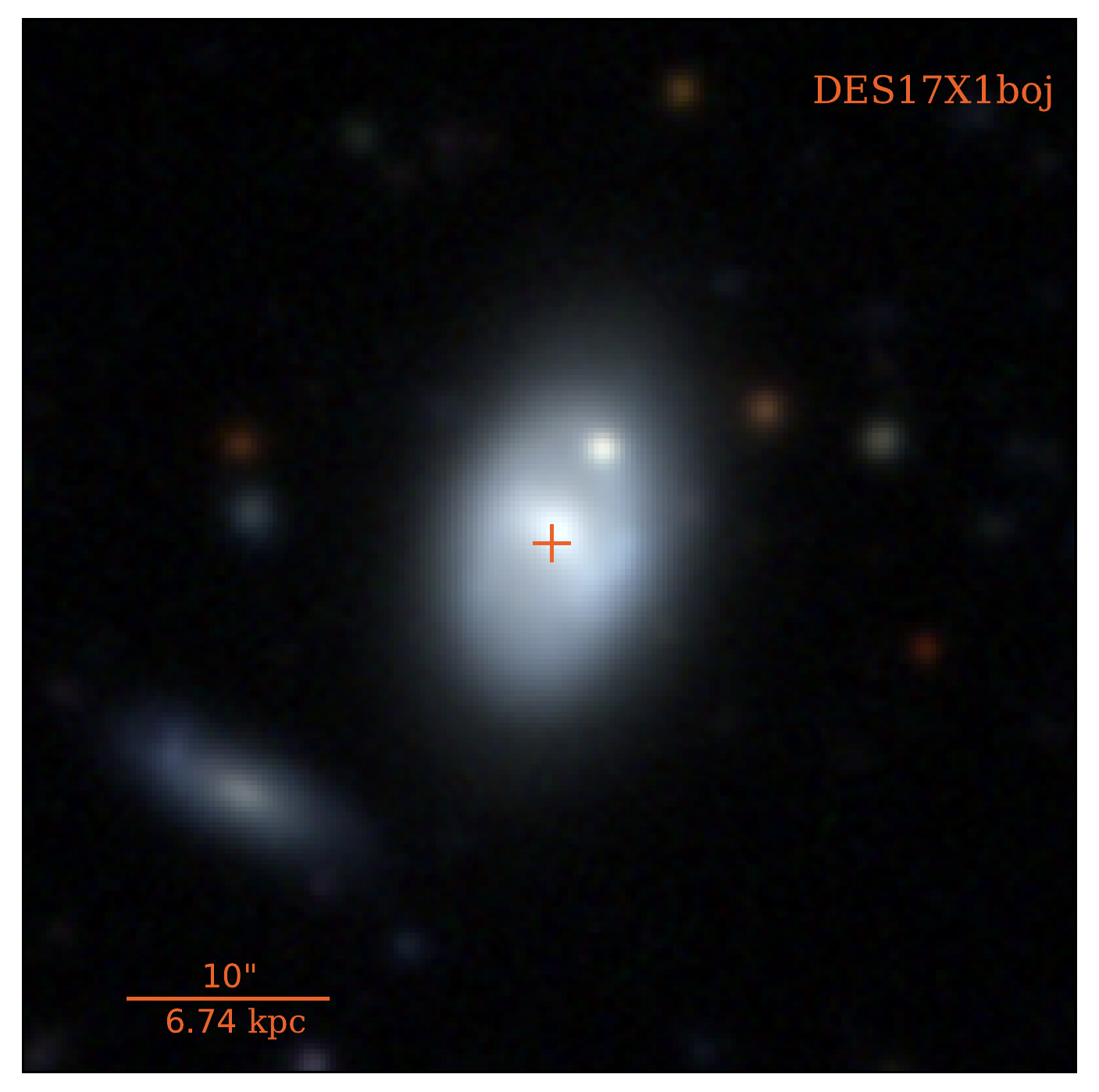}
    \end{subfigure}%
    ~ 
    \begin{subfigure}[b]{0.45\textwidth}
        \centering
        \includegraphics[width=\textwidth]{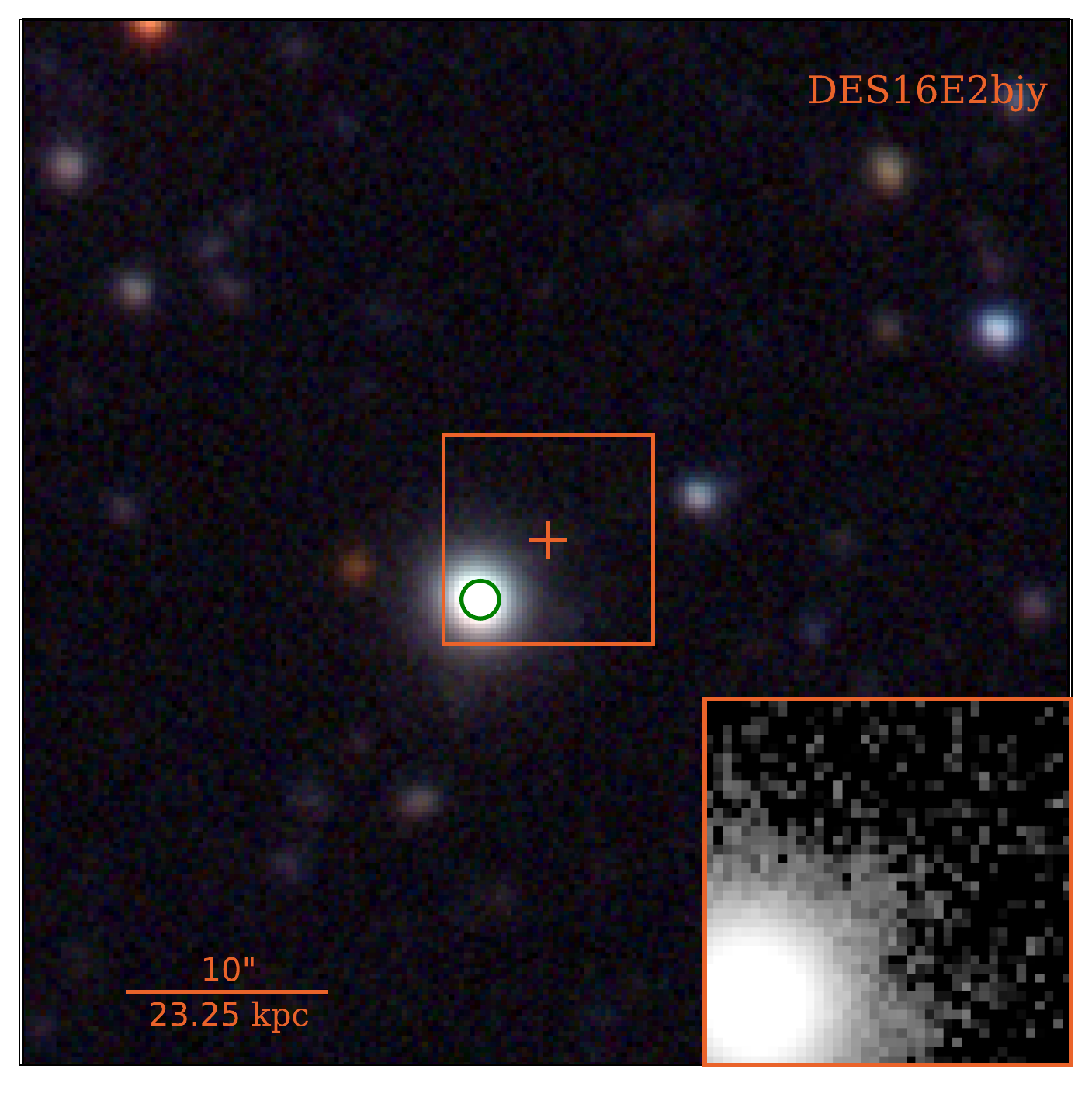}
    \end{subfigure}
    \caption{Host environments of DES17X1boj and DES16E2bjy. Transient locations are marked with red crosses and
    the assumed host of DES16E2bjy is marked with green circle. The host of DES17X1boj is at $z=0.0338$ and DES16E2bjy at $z=0.1305$ For DES16E2bjy we also show a zoom-in of the $r$-band image around the transient in the bottom-right corner, revealing the presence of light presumably of the host galaxy at the location.}
    \label{fig:stamps}
\end{figure*}

The Dark Energy Survey Supernova (DES-SN) Programme has proven to be an excellent laboratory for discovering unusual types of events with its deep ($\sim 24$ mag per visit) multicolor $griz$ photometry. It has, for example, already probed the diversity of a sample of SLSNe \citep{Angus2018} with objects up to $z\approx2$ \citep[DES16C2nm; ][]{Smith2017} and unveiled a sample of rapidly evolving transients \citep{Pursiainen2018}. In this paper we present two additional peculiar transients discovered by DES-SN, DES17X1boj and DES16E2bjy, with highly atypical double-peaked light curves nearly identical to each other. 

Light curves exhibiting phases of rebrightening are not uncommon amongst extragalactic transients. Type Ia SNe have a characteristic secondary peak in the redder bands due to Fe III recombination \citep{Kasen2006}, and a fraction of core-collapse SNe (CCSNe) and SLSNe also show similar behaviour. For example, several type IIb SNe \citep[e.g. SN1993J; ][]{Wheeler1993}, type Ic iPTF14gqr \citep{De2018a} and several SLSNe  \citep[e.g. SN2006oz; ][]{Leloudas2012} all have short precursory bumps often attributed to shock cooling in extended material surrounding the SN. Additionally, several type IIP SNe have shown shallow rising during their plateau phase \citep[e.g. SN2009N;][]{Takats2014}. However, none of these double-peaked SNe have similar light curves to the DES-SN transients, as they have differences in the duration and the strength of the two peaks. 

Here we present an analysis of the photometric and spectroscopic data for the DES-SN photometric twins, DES17X1boj and DES16E2bjy.  While they share nearly identical light curve evolution, several other observable features are clearly different.  The transients differ by nearly three magnitudes in brightness with DES17X1boj peaking at $M_\mathrm{r}=-15.4$ and DES16E2bjy reaching $M_\mathrm{r}=-17.9$. They are also found in different host environments. DES17X1boj's position is consistent with the nucleus of a star-forming spiral galaxy, while DES16E2bjy appeared on the outskirts of a passive galaxy, with the only common factor being a presence of old stellar populations in both environments. While the high-velocity ($9400$ km/s) CaII absorption feature in the spectrum of DES17X1boj and the photospheric expansion velocity of DES16E2bjy ($v\approx4800$ km/s) strongly indicate an explosive or eruptive origin for these transients, the exact physical scenario remains unknown. Due to the several different observed features it is also entirely possible that the transients do not share a similar origin, despite similar light curve evolution. 

In Section \ref{sec:observations} we provide description of the observational data used in this paper and in Sections \ref{sec:analysis} and  \ref{sec:hostgal} we present the analysis of the transients and their host galaxies. Finally in Section \ref{sec:discussions} we discuss the physical scenarios that could explain the peculiar observed features of the transients and summarise our findings. Throughout this paper we assume a flat $\Lambda$CDM cosmology with $\Omega_\mathrm{M}=0.3$ and H$_\mathrm{0}$ = 70 km s$^{-1}$ Mpc$^{-1}$. 

\section{Observations}
\label{sec:observations}

\begin{figure*}
    \centering
    \begin{subfigure}[b]{0.4\textwidth}
        \centering
        \includegraphics[width=\textwidth]{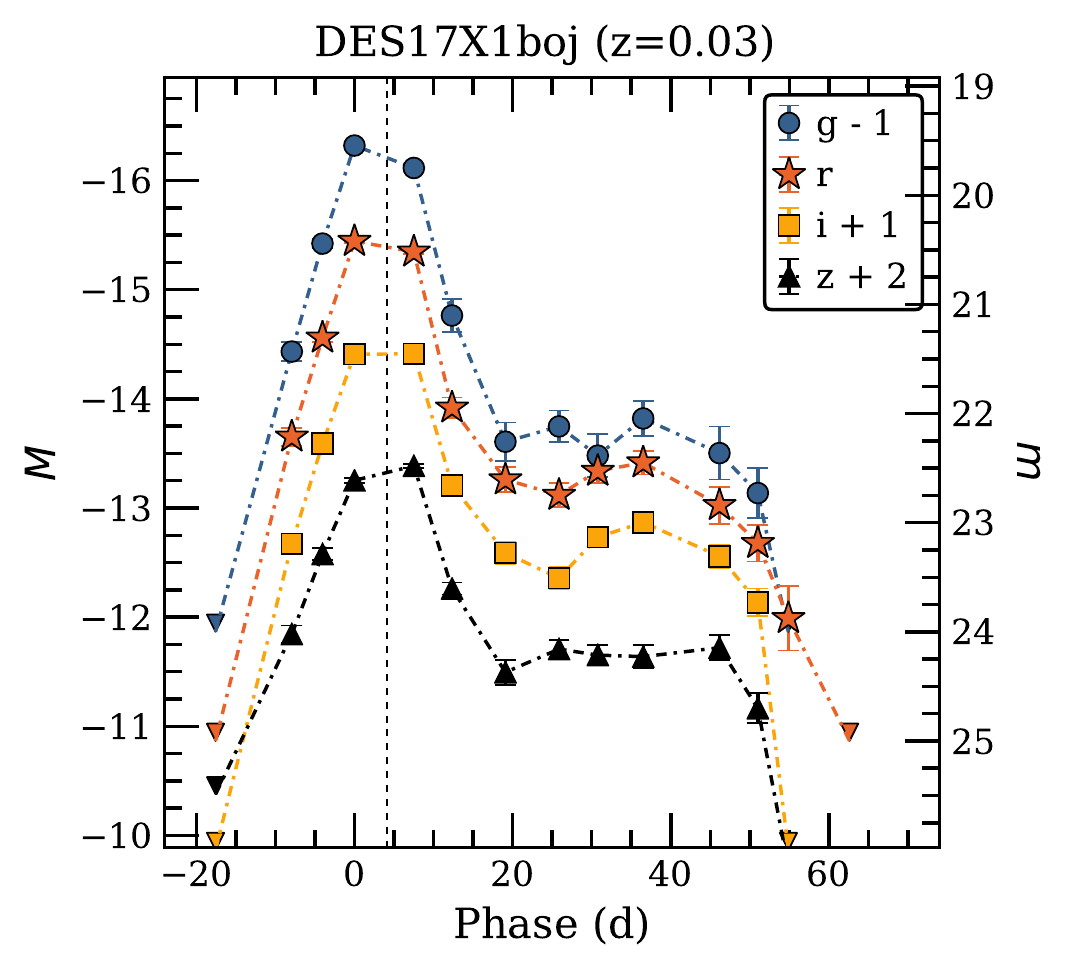}
    \end{subfigure}%
    ~ 
    \begin{subfigure}[b]{0.4\textwidth}
        \centering
        \includegraphics[width=\textwidth]{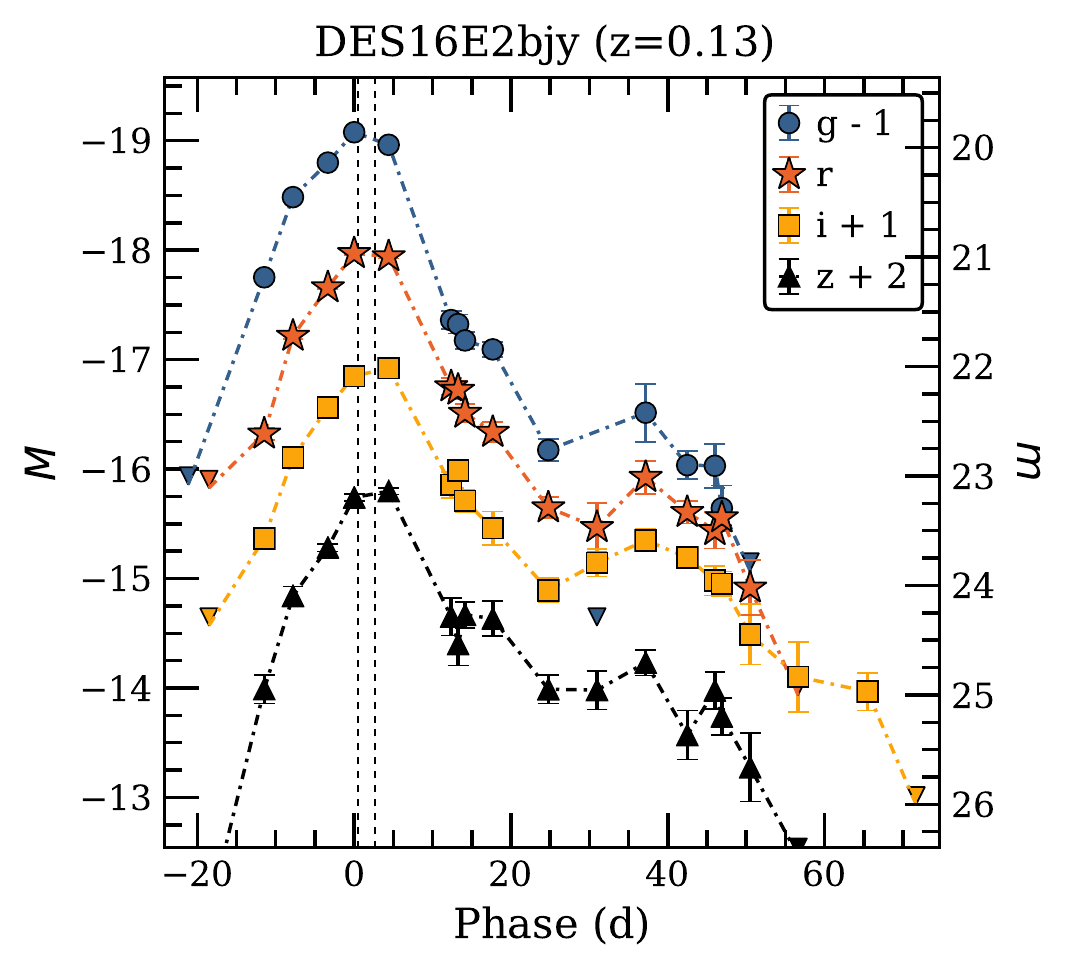}
    \end{subfigure}
    \caption{Rest frame DES-SN $g,r,i,z$ light curves of DES17X1boj and DES16E2bjy, with respect to maximum brightness in $g$ band. Photometric bands are offset for clarity. Downward-facing triangles represent upper limits. The phases of spectroscopic observations have been marked with dashed vertical lines.}
    \label{fig:lcs}
\end{figure*} 

DES17X1boj was first detected using the \texttt{DiffImg} pipeline \citep{Kessler2015} at a signal-to-noise $> 5$ by DES-SN in Dark Energy Camera \citep*[DECam; ][]{Flaugher2015} $griz$ images taken on 9th of October 2017 at $\alpha=02^\mathrm{h}19^\mathrm{m}53.80^\mathrm{s}$, $\delta=-04^\circ57\arcmin10.4\arcsec$ (J2000) with a previous non-detection on the 29th of September. It reached a maximum brightness of $m_\mathrm{r}=20.52$ ($M_\mathrm{r}=-15.35$) on the 17th of October and was detected for the last time on 13th of December the same year. In the beginning of the light curve, DES17X1boj was identified as a potential candidate for a rapidly evolving transient in a live selection of targets during year five of DES-SN, where we visually inspected all new transients with relatively blue colors ($g-r < 0.2$). Promptly after the discovery it was reclassified as another type of peculiar object due to the double-peaked light curve. The transient was observed spectroscopically for 2400 s on the 22nd of October as part of the OzDES programme (\citealt{Yuan2015}; \citealt{Childress2017}; Tucker et al. \textit{in prep}) and the resulting data was reduced based on the pipeline described in \citet{Yuan2015}. OzDES uses the 3.9~m Anglo-Australian Telescope (AAT) in Australia with the AAOmega spectrograph and with the 2dF robotic fibre positioner. The resulting spectra has an average resolution of $R\sim1400$, with 1 Å/pix dispersion in the blue arm and 1.6 Å/pix in the red arm \citep{Yuan2015}. As the underlying galaxy is $1.0-1.5$ magnitudes brighter than DES17X1boj at peak brightness in all bands, the spectrum is dominated by host galaxy features. The host of DES17X1boj was previously observed with the AAT under the Galaxy and Mass Assembly \citep[GAMA; ][]{Driver2009, Baldry2018} programme for 7200 s in November 2010. 

DES16E2bjy was recovered with an archival search for transients that exhibited similar light curve evolution to DES17X1boj that was performed after the end of DES-SN operations (see Section \ref{subsec:gp}). The transient was first detected by DES-SN on the 21st of September 2016 at $\alpha=00^\mathrm{h}33^\mathrm{m}19.34^\mathrm{s}$, $\delta=-44^\circ20\arcmin19.6\arcsec$ (J2000), with a previous non-detection on the 13th of September. It reached maximum brightness of $m_\mathrm{r}=20.99$ ($M_\mathrm{r}=-17.94$) on the 4th of October and was detected for the last time on the 17th of December. DES16E2bjy was spectroscopically observed for 2400 s on the 5th of October by OzDES, and for 2800 s on the 7th of October 2016 with Robert Stobie Spectrograph \citep[RSS; ][]{Nordsieck2001} on 10 meter-class South African Large Telescope (SALT) as it was a high probability candidate for being a type Ia SN. The SALT spectrum has a resolution of $R\sim350$ and it was reduced using the \textsc{PySALT}\footnote{http://pysalt.salt.ac.za/} pipeline \citep{Crawford2010}. All presented light curves have been corrected for the Milky Way extinction using color excesses from \citet{Schlafly2011}. Further details on DES-SN difference imaging photometry can be found in \citet{Kessler2015} and on DES-SN spectroscopic follow up programs in \citet{DAndrea2018}. The photometric properties are presented in Section \ref{subsec:phot} and spectroscopic properties in Section \ref{subsec:spectra}. 

The transient environments are presented in Figure \ref{fig:stamps}, with stamps created using the optimised DES-SN deep stacks \citep{Wiseman2020}. The location of DES17X1boj is consistent with the nucleus of a spiral galaxy \citep[LEDA 1051269, ][]{Brouty2003} at $z=0.0338$, while DES16E2bjy appears to be associated with a red elliptical galaxy at $z=0.1305$ with a separation of $\approx 5\arcsec$ (11.6 kpc). The host galaxy redshifts were obtained by the OzDES programme and all presented absolute magnitudes have been calculated based on them. The host galaxy properties are discussed in Section \ref{sec:hostgal}. 

\begin{figure*}
    \centering
    \includegraphics[width=0.98\textwidth]{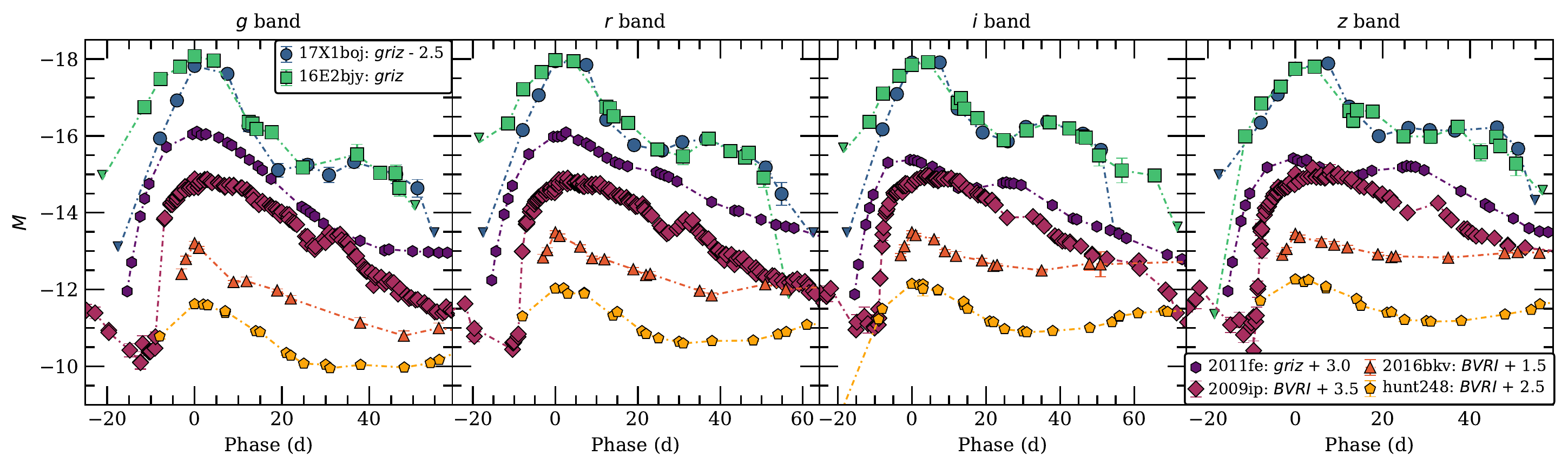}
    \caption{Light curves of DES16E2bjy and DES17X1boj, with respect to peak in $g$ band, scaled by 2.5 magnitudes to match in maximum brightness. Type Ia SN2011fe \citep{Richmond2012, Munari2013, Tsvetkov2013}, SN impostor SN2009ip \citep[during the event in 2012; ][]{Fraser2013,Mauerhan2013, Pastorello2013, Brown2014}, low-luminosity type IIP SN2016bkv \citep{Hosseinzadeh2018,Nakaoka2018} and triple-peaked transient event SNhunt248 \citep{Kankare2015, Mauerhan2015}, proposed to be a Luminous Red Novae \citep[LRN, ][]{Mauerhan2018}, have been plotted for comparison. The light curves of literature events were collected from \textit{The Open Supernova Catalog} \citep{Guillochon2016}, apart from SN2011fe for which the light curve was created based on BVRI light curves from the above sources which were $K$-corrected \citep{Nugent2002} using a spectrophotometric templates from \citet{Hsiao2007}. Note that offsets have been added to literature events for visual clarity.}
    \label{fig:lcs_comp}
\end{figure*}

\subsection{Sample Selection}
\label{subsec:gp}

After DES17X1boj was discovered, an archival search was performed on the full five years of DES-SN data to search for transients with similar light curve evolution. For this purpose we generated interpolated light curves of all $\sim$30000 DES-SN transients using Gaussian Processes (GP) with a similar approach to \citet{Angus2018} and \citet{Inserra2018}. GP use the uncertainties of the flux measurements to determine how strongly sequential data points are correlated, and then interpolate the light curves based on a predefined function (``kernel'') that defines how strongly the light curve may vary in time. In this paper we use the squared exponential model for the kernel with timescale of variation which was determined using gradient based optimisation. The light curves were created with 0.5 d cadence, so that every GP epoch had a flux value in all four bands. Every year and band was interpolated independently with the first epoch being at the time of the first observation of the year regardless of the band. The interpolation was performed with the \texttt{python} package \texttt{george} \citep{Ambikasaran2014}. 
Using the GP light curves we performed a systematic search for DES-SN transients with similar light curve evolution to DES17X1boj. In detail, the GP light curves were used to estimate the peak times and brightnesses of each named DES-SN transient so that we could easily find events fulfilling our search criteria. 

We searched for all unclassified DES-SN transients for which a secondary peak of a `boj-like' transient would have been seen (at least 24 mags). This requirement corresponds to brightness of the primary peak of at least 22 mags. However, we also constrained the the total observable volume by requiring the transient or its associated host galaxy to have a redshift $z\leq0.2$. At this redshift the peak brightness of a transient observed at 22 mag would correspond to $M=-18$. With the given criteria, DES17X1boj ($M_\mathrm{r}=-15.4$) could have been recovered only up to $z=0.07$. However, the search was designed to maximise our chance of discovering `boj-like' transients while also allowing them to be brighter than DES17X1boj. 

In total 225 DES-SN transients passed the criteria. Light curves of these transients were visually inspected and only DES16E2bjy was found to have a similar double-peaked light curve evolution as DES17X1boj. Most of the other inspected transients had light curves similar to the traditional SN types and the few remaining transients were found to be spurious detections.

\section{Analysis}
\label{sec:analysis}
\subsection{Photometric Properties}
\label{subsec:phot}

As shown in Figure \ref{fig:lcs}, DES17X1boj and DES16E2bjy share very similar light curve evolution. Both rise to maximum brightness, as measured from photometry in $g$ band, in $15 - 20$ d and decline rapidly by 2 - 3 magnitudes in $\sim20$ d before plateauing and rising to the second peak-like feature. After the secondary peak both transients fade quickly below the detection limit.  However, despite very similar light curves, the peak luminosities are very different: while DES17X1boj is $M_\mathrm{r}=-15.4$ at observed maximum brightness, DES16E2bjy reaches $M_\mathrm{r}=-17.9$. The presented light curves have not been $K$-corrected due to unknown Spectral Energy Distributions (SEDs) during the secondary peak. However, we estimated the significance of $K$-correction at maximum brightness based on blackbody fits to the photometry (see Appendix \ref{app:bb_fits}). The correction for DES17X1boj is less than 0.1 mag and for DES16E2bjy 0.2 to 0.3 mag so that each band becomes fainter.

The similarity of the light curves is further emphasised in Figure \ref{fig:lcs_comp}, where a scaling of 2.5 mags (factor of ten in brightness) has been added to DES17X1boj to match the peak luminosities. The scaled light curves are nearly identical with the only significant difference between the two transients being the rise to maximum brightness which appears to be longer for DES16E2bjy especially in the bluer bands. The rise also appears to have two phases; a fast rise until $\approx10$ d before peak and a shallower, nearly linear rise to maximum brightness. 

We have also plotted light curves of both normal and peculiar transients from the literature in Figure \ref{fig:lcs_comp} to demonstrate that the light curve evolution of DES17X1boj and DES16E2bjy is highly atypical. The transients have been selected as they evolve in similar timescales as the DES transients and show phases of rebrightening at similar epochs. Type Ia SNe (here illustrated with SN2011fe) always have a secondary peak but it is prominent only in the redder bands, while DES17X1boj and DES16E2bjy exhibit a strong secondary peak in all four optical bands. Additionally the shape of the secondary peaks seen in $i$ and $z$ bands is very different from the DES-SN transients. Some type IIP SNe, such as SN2016bkv, show shallow rising during their plateau phase, but on much longer timescales and without significant changes in brightness. Even other peculiar transients with several phases of rebrightening, such as impostor SN2009ip (during the event in 2012) and Luminous Red Nova (LRN) SNhunt248, do not show as extreme variation as seen in the light curves of DES17X1boj and DES16E2bjy. 

\begin{figure}
    \centering
    \includegraphics[width=0.47\textwidth]{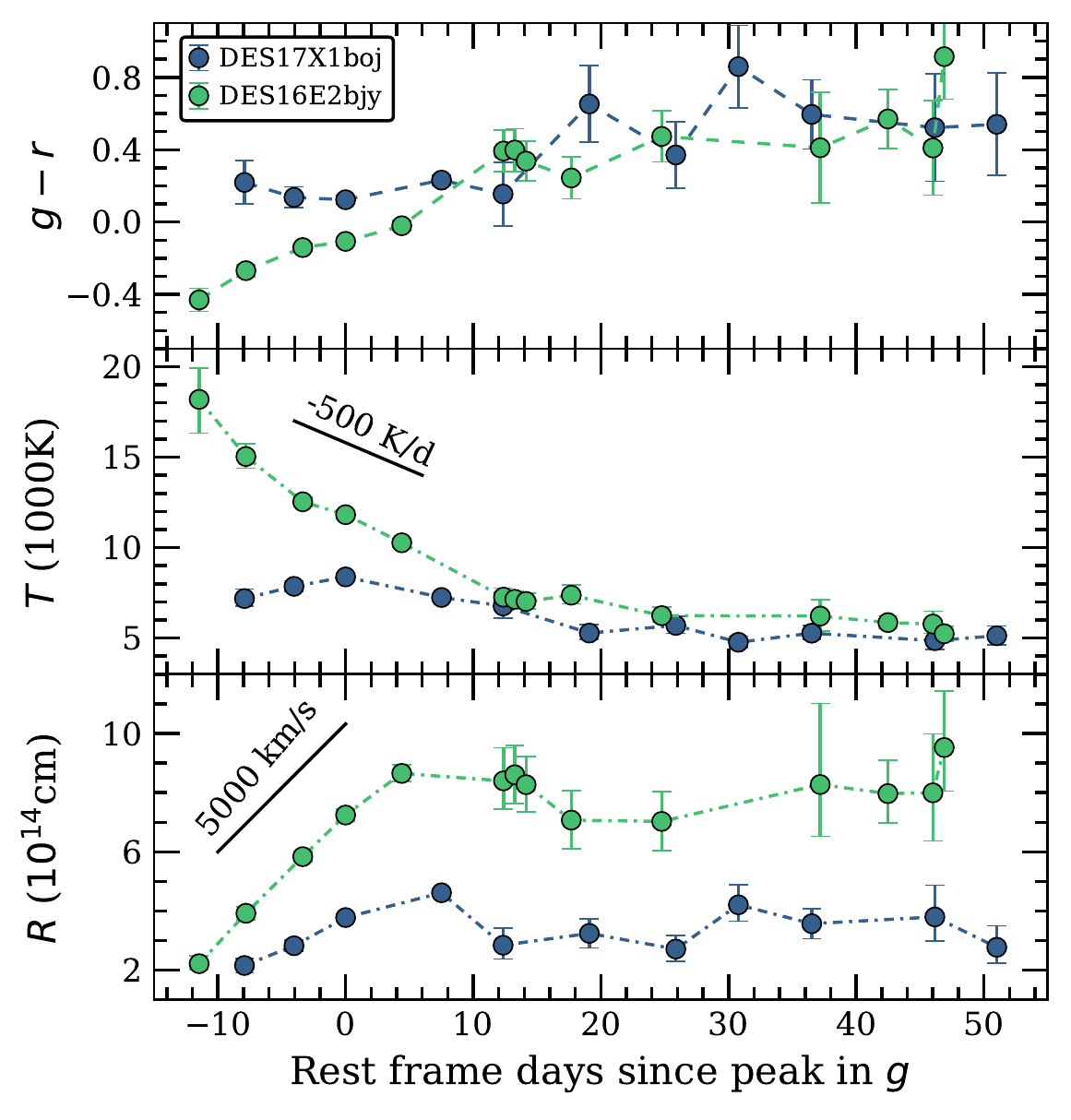}
    \caption{Observed $g-r$ colour (top) with temperature (middle) and radius evolution (bottom) based on blackbody fits to photometry. Errors for temperature and radius are given in $1\sigma$ confidence. Reference curves for velocity and change of temperature have been added for clarity.}
    \label{fig:TR_evo}
\end{figure}

We performed blackbody fits to each epoch of the four-band photometry for both of the transients and the fits are shown in Appendix \ref{app:bb_fits}, Figures \ref{fig:bb_boj} and \ref{fig:bb_bjy}, for DES17X1boj and DES16E2bjy respectively. The photometric data is well described with a blackbody at least until $\sim10$ d after initial peak. At later times the blackbody fits are less constrained due to fading targets. The blackbody temperature and radial evolution are shown in Figure \ref{fig:TR_evo} with the observed $g-r$ colour evolution. DES16E2bjy is clearly bluer and therefore hotter in the beginning of the light curve, but after roughly $+10$~d the transients share similar colour and temperature evolution. However, while DES162E2bjy starts with a high temperature of $T \approx18000$ K and quickly cools down, the temperature of DES17X1boj increases slightly until the peak brightness, reaching a temperature of $\approx8000$ K. Afterwards the temperature gradually decreases and there seems to be no evidence of significant colour evolution. On the other hand, the radii of the transients evolve similarly to each other. First the radius increases nearly linearly until $\approx+10$ d, after which the radius either stays constant or decreases slightly. The photospheric expansion velocities based on the blackbody fits up to $+10$ d are $v\approx1800$ km/s for DES17X1boj and $v\approx4800$ km/s for DES16E2bjy. Based on the sharp transition in the evolution of radii at roughly $+10$ d, it is likely that SEDs are not described by blackbody emission alone after this epoch (see Appendix \ref{app:bb_fits}).

\begin{figure}
    \centering
    \includegraphics[width=0.47\textwidth]{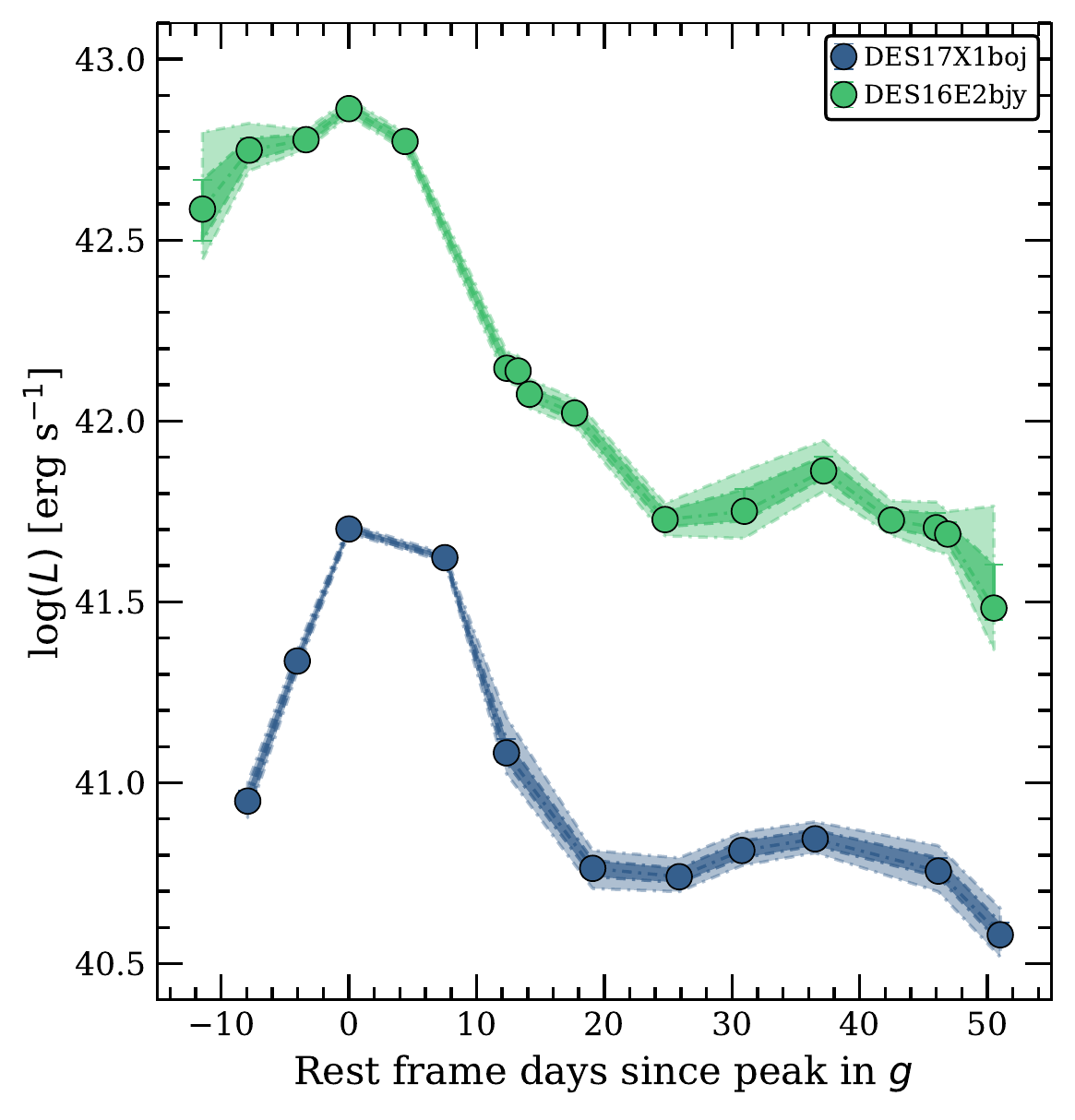}
    \caption{Pseudo-bolometric light curves of DES17X1boj and DES16E2bjy. Light curves are constructed using the blackbody fits and the shaded regions refer to 1$\sigma$ and 2$\sigma$ confidence levels.}
    \label{fig:lc_bol}
\end{figure} 

\begin{figure}
    \centering
    \includegraphics[width=0.47\textwidth]{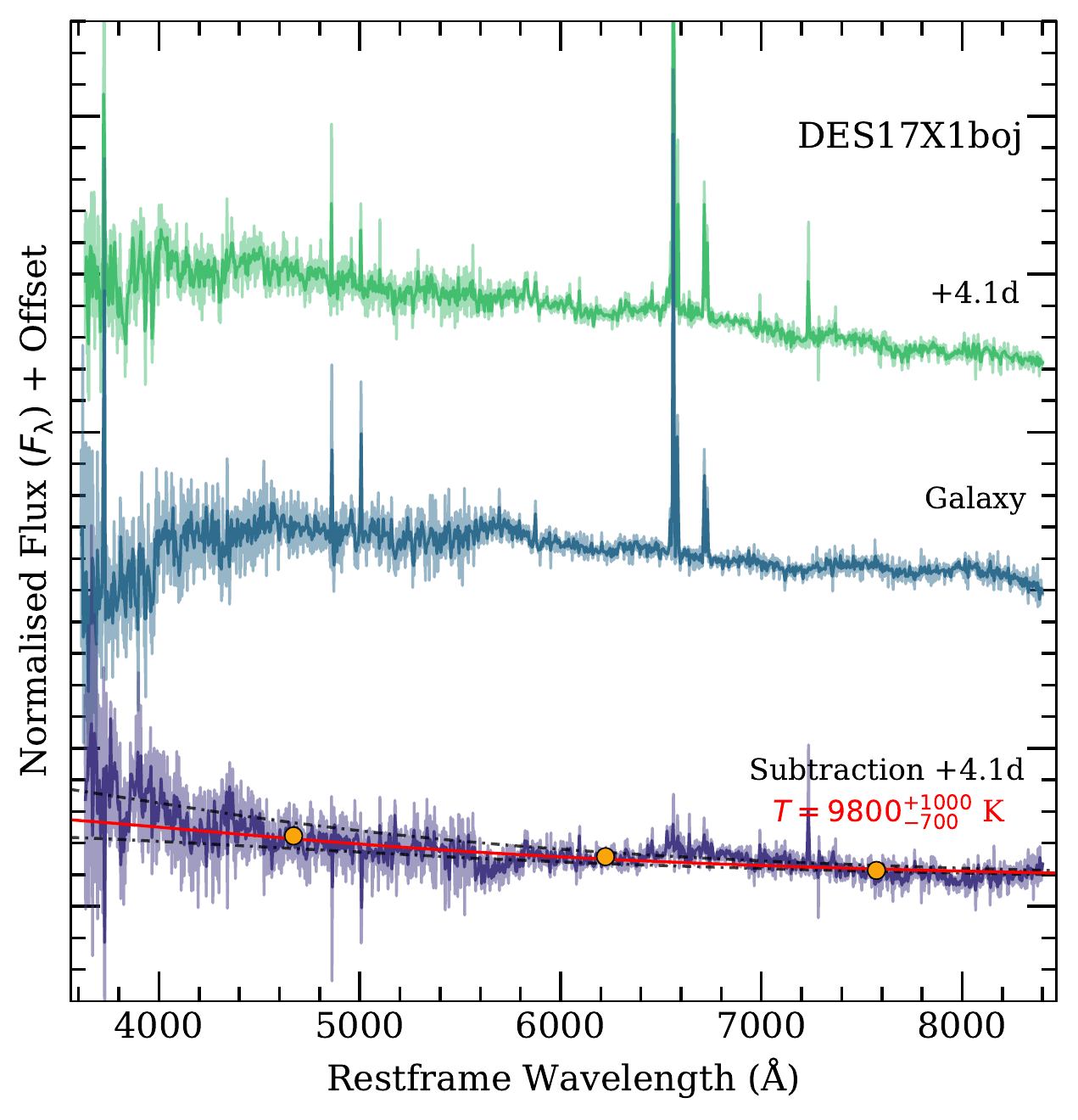}
    \caption{AAT spectrum of DES17X1boj, the host galaxy spectrum and the resulting host galaxy subtraction (lighter shade) and the corresponding spectra binned by factor of five (darker shade). The transient spectrum is taken a few days after the observed peak brightness and holds a considerable amount of transient light. The subtraction shows a blue continuum with blue-shifted CaII absorption feature at $3830$ Å. The jump at 5600Å in the subtraction is caused by a dichroic issue in the galaxy spectrum. The red line is the best-fitting blackbody to the subtraction with black dashed lines corresponding to the given 1$\sigma$ errors on the temperature. The orange circles are photometric $gri$ data taken within a few days of the spectrum.}
    \label{fig:spectra_boj}
\end{figure} 

In Figure \ref{fig:lc_bol} we present pseudo-bolometric light curves of DES17X1boj and DES16E2bjy constructed using the blackbody fits (see Appendix \ref{app:bb_fits}). The double-peaked shape of both light curves is clearly visible in the bolometric light curves. Interestingly the two-phased rise of the initial peak of DES16E2bjy seen in the $griz$ light curves (see Figure \ref{fig:lcs}) appears to be slightly more pronounced in the bolometric light curve.

\subsection{Spectroscopic properties}
\label{subsec:spectra}

\begin{figure}
    \centering
    \includegraphics[width=0.47\textwidth]{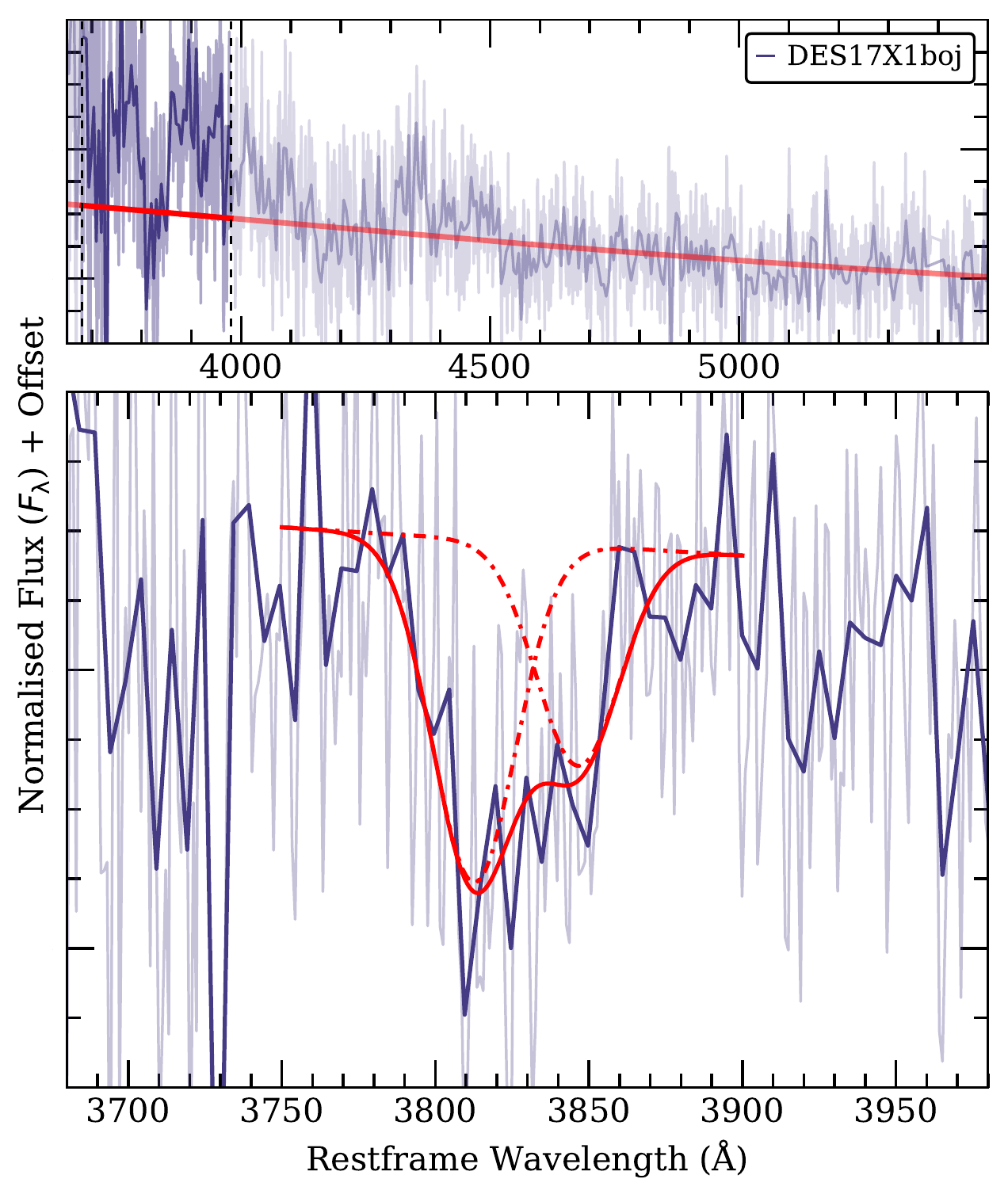}
    \caption{ Spectrum of DES17X1boj highlighting the blueshifted H and K absorption with the best-fitting blackbody presented in Figure \ref{fig:spectra_boj} (top) and the absorption feature with best-fitting Gaussian profiles given in red.  Lines were fixed to be at $3813$ Å and $3846$ Å (corresponding to ejecta velocity of $v\approx9400$ km/s). Width of the lines was found to be $v_\mathrm{FWHM}=2500^{+1600}_{-600}$ km/s. Note that the width of the lines was forced to be the same when fitting. The darker shade represents the binned data as per Figure \ref{fig:spectra_boj}.}
    \label{fig:boj_caii}
\end{figure} 

\begin{figure}
    \centering
    \includegraphics[width=0.47\textwidth]{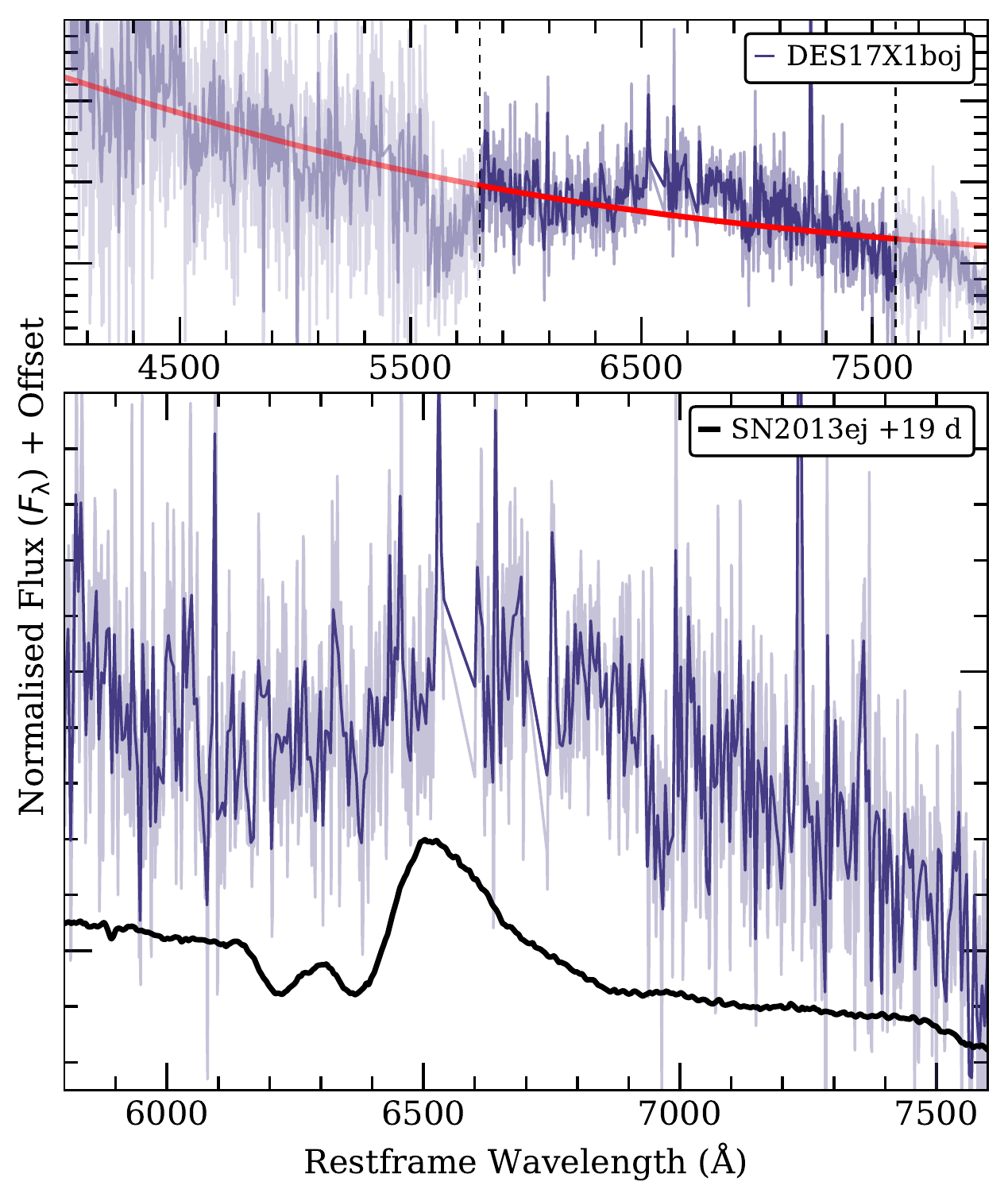}
    \caption{Spectrum of DES17X1boj highlighting the excess emission around the H$\alpha$ line on top and the the H$\alpha$ region of DES17X1boj in comparison with type II SN2013ej around peak brightness, 19 days after the explosion \citep{Dhungana2016} on bottom. Solid red line is the best-fitting blackbody presented in Figure \ref{fig:spectra_boj}. Note that residuals of narrow lines left from the host galaxy subtraction have been masked out. The darker shade represents the binned data as per Figure \ref{fig:spectra_boj}.} 
    \label{fig:boj_halpha}
\end{figure} 

\begin{figure}
    \centering
    \includegraphics[width=0.47\textwidth]{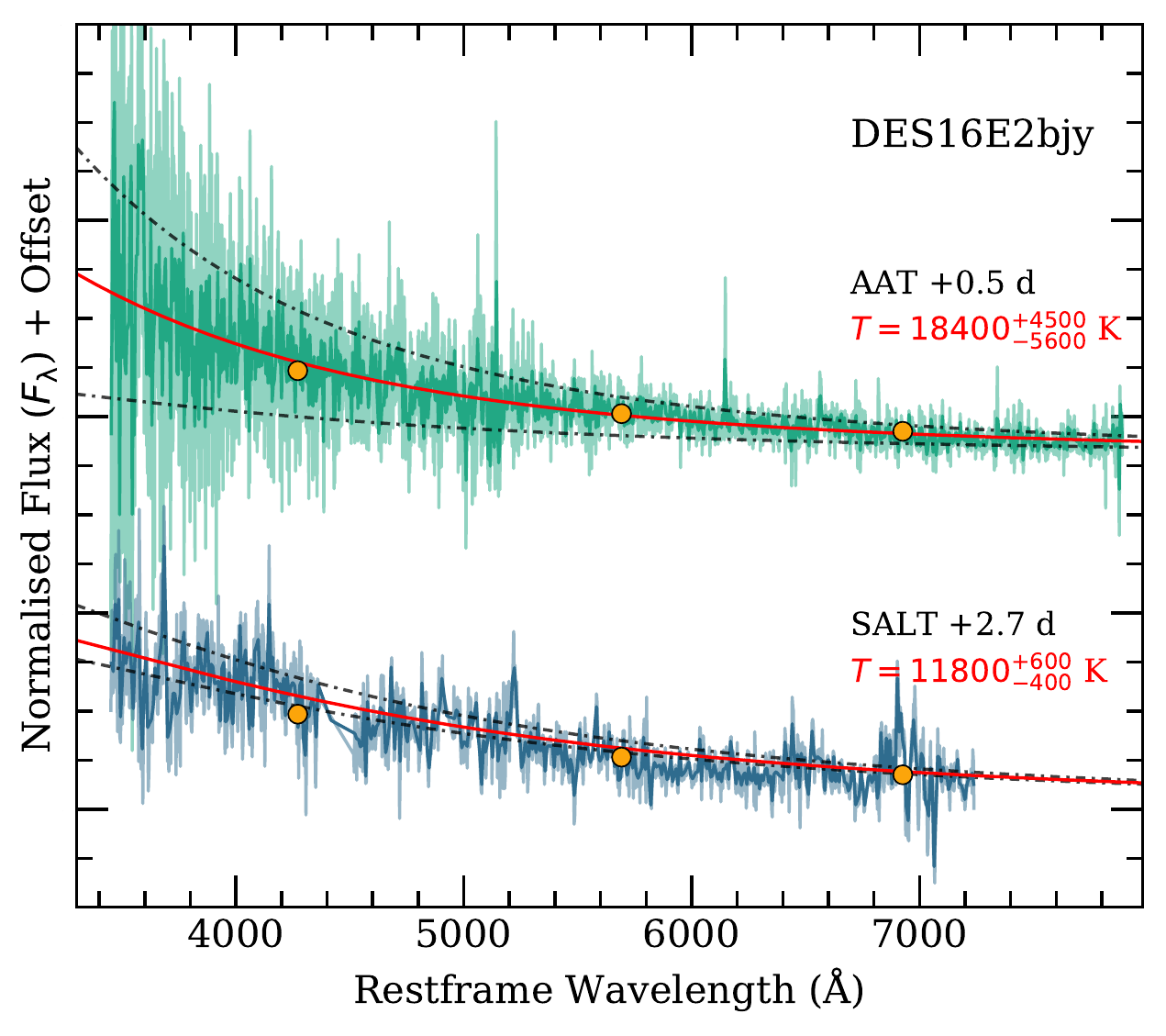}
    \caption{AAT and SALT spectra of DES16E2bjy close to peak brightness. The red line is the best-fitting blackbody to the spectra with the black dashed lines corresponding to the given 1$\sigma$ errors on the temperature. and the orange circles are photometric data points taken within few days of the corresponding spectrum. Note that some strong residuals from sky subtraction have been removed for visual clarity and there is a chip gap in the SALT spectrum around 4400 Å.}
    \label{fig:spectra_bjy}
\end{figure}

The AAT spectrum of DES17X1boj is presented in Figure \ref{fig:spectra_boj}, with the host galaxy spectrum and resulting host galaxy subtraction. The spectrum was obtained four days after the observed maximum brightness, but even at this epoch the host galaxy was $1.0-1.5$ mag brighter than the transient in all bands and thus the spectrum was dominated by host galaxy features. After the host galaxy subtraction, the spectrum shows a blue continuum best fit with a blackbody of $T=9800_{-700}^{+1000}$ K which is slightly higher than the temperature based on the photometry ($T\approx8000$ K, see Figure \ref{fig:TR_evo}). The only significant line feature clearly visible in the subtraction is a relatively broad absorption at $\lambda\approx3830$ Å (measured at the middle of the absorption feature). This feature resembles blueshifted calcium H\&K absorption often seen in various types of SNe (see e.g. type Ia SN2011fe \citep{Silverman2015}, type IIb SN2011dh \citep{Arcavi2011} and Ca-rich transient SN2005E \citep{Perets2010}). Based on the wavelength at the middle of the feature, the corresponding ejecta velocity would be $v\approx9400$ km/s. The feature has been plotted in Figure \ref{fig:boj_caii}, where we also show the best-fitting Gaussian absorption lines. The fitted Gaussian profiles were forced to be at $3813$ Å and $3846$ Å (corresponding to ejecta velocity of $v=9400$ km/s) and to have the same width that was free to vary. The value was found to be $v_\mathrm{FWHM}=2500^{+1600}_{-600}$ km/s, which is significantly higher than the resolution of the instrument $\Delta\lambda \sim 250$ km/s at H$\mathrm{\alpha}$. All given errors were estimated using a Monte Carlo approach with 500 realisations and are shown in 1$\sigma$ confidence.

The high velocity of this blueshifted CaII absorption is in good agreement with what is seen in SNe and could thus suggest that DES17X1boj is a SN of some type. For instance, in the spectra of type IIb SN2011dh the feature occurs at higher velocity during the rise but it is found at $v\approx10000$ km/s at the time of peak brightness \citep{Sahu2013}. On the other hand, the photospheric expansion velocity of this SN is $v\approx 3000$ km/s estimated based on the evolution of the blackbody radius presented in \citet{Ergon2014}. While the value is higher than what was found for DES17X1boj, it demonstrates that a large velocity difference between the photosphere and the absorbing material can occur and has already been observed. \citet{Ergon2014} also estimated photospheric temperature evolution with a peak temperature of $T\approx8000 - 9000$ K, similar to what is seen in DES17X1boj. Despite the discussed similarities, the light curve evolution of SN2011dh is smoother and slower than seen in DES17X1boj apart from the short duration pre-peak often seen in type IIb SNe \citep{Arcavi2011}.

In Figure \ref{fig:spectra_boj} there seems to be small amount of excess emission above the plotted blackbody fit around $6600$ Å. To investigate this in detail, we show the spectrum of DES17X1boj around the H$\alpha$ line in Figure \ref{fig:boj_halpha} in comparison with type II SN2013ej 19 d after explosion (\citealt{Dhungana2016}, see also e.g. \citealt{Valenti2014}, \citealt{Bose2015} and \citealt{Huang2015}). This SN was selected as it has a similar rise time to DES17X1boj in the $r$ band and the spectrum was taken at comparable epoch around the peak brightness. There appears to be excess emission that slightly resembles a broad H$\alpha$ profile which would indicate that DES17X1boj is a type II SN. While the similarity appears to be reasonable, the high level of noise in the spectrum makes it difficult to say anything definite about the feature. 

The AAT and SALT spectra of DES16E2bjy, taken shortly after peak brightness, are presented in Figure \ref{fig:spectra_bjy}. Neither of the spectra show other features than underlying blue continuum which is best characterised with a blackbody. A blackbody fit to the AAT spectrum taken at $+0.5$ d after peak brightness in $g$ band results in a temperature of $T=18400_{-5600}^{+4500}$ K, which is higher than that based on the blackbody fits to photometry ($T\approx12500$ K, see Figure \ref{fig:TR_evo}) due to the high level of noise in the blue part. The SALT spectrum at $+2.7$ d shows a blackbody with $T=11800_{-400}^{+600}$ K, consistent with the value based on photometry. All presented spectra have been flux calibrated using the DES-SN \textit{griz} photometry of the transients at the date closest to the epoch of spectra (the maximum difference being 3 days). For DES17X1boj the aperture photometry of the host from the DES Science Verification \citep[SV; ][]{Bonnett2016, Rykoff2016} data was also applied. In the case of DES16E2bjy, the host galaxy brightness was estimated to be $m \approx 24.8 - 26.0$ (depending on the band) in 2$\arcsec$ aperture. This is significantly fainter than the transient at peak brightness ($m \approx 21$) when both of the spectra were taken and thus the contribution of the host galaxy is assumed to be negligible in the spectra.

\section{Host Galaxy Properties}
\label{sec:hostgal}

\begin{figure}
    \centering
    \includegraphics[width=0.48\textwidth]{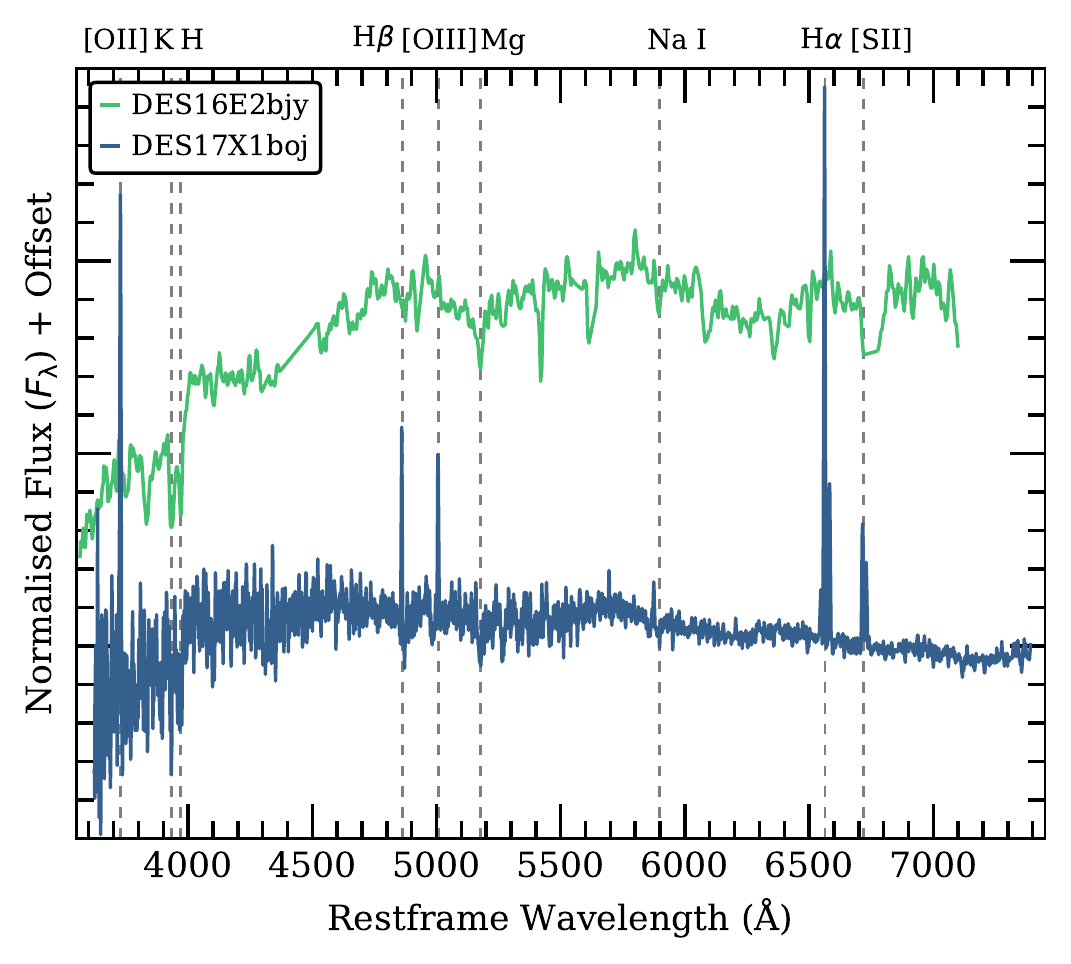}
    \caption{Host spectra of DES17X1boj and DES16E2bjy. Some common absorption and emission lines have been marked with dashed lines.}
    \label{fig:host_spectra}
\end{figure}

DES17X1boj and DES16E2bjy seem to have occurred in different host environments (see Figure \ref{fig:stamps}), and to further demonstrate this we have plotted the host spectra in Figure \ref{fig:host_spectra}. The host of DES17X1boj was observed with the AAT under the GAMA programme with a 2$\arcsec$ fibre, positioned at the centre of the host galaxy where the transient also occurred. As DES16E2bjy was separated by 5$\arcsec$ from its host, we obtained the host spectrum at the core from the long-slit transient spectrum observed with SALT. Based on the Figure \ref{fig:host_spectra}, it is clear that the hosts of these two transients are different: DES17X1boj happened in a star-forming galaxy characterised by numerous strong nebular emissino lines, while the host of DES16E2bjy is a passive galaxy with no obvious signatures of recent star formation in the optical spectrum (e.g. H$\alpha$, [OIII], [NII]). 

\begin{figure}
    \centering
    \includegraphics[width=0.48\textwidth]{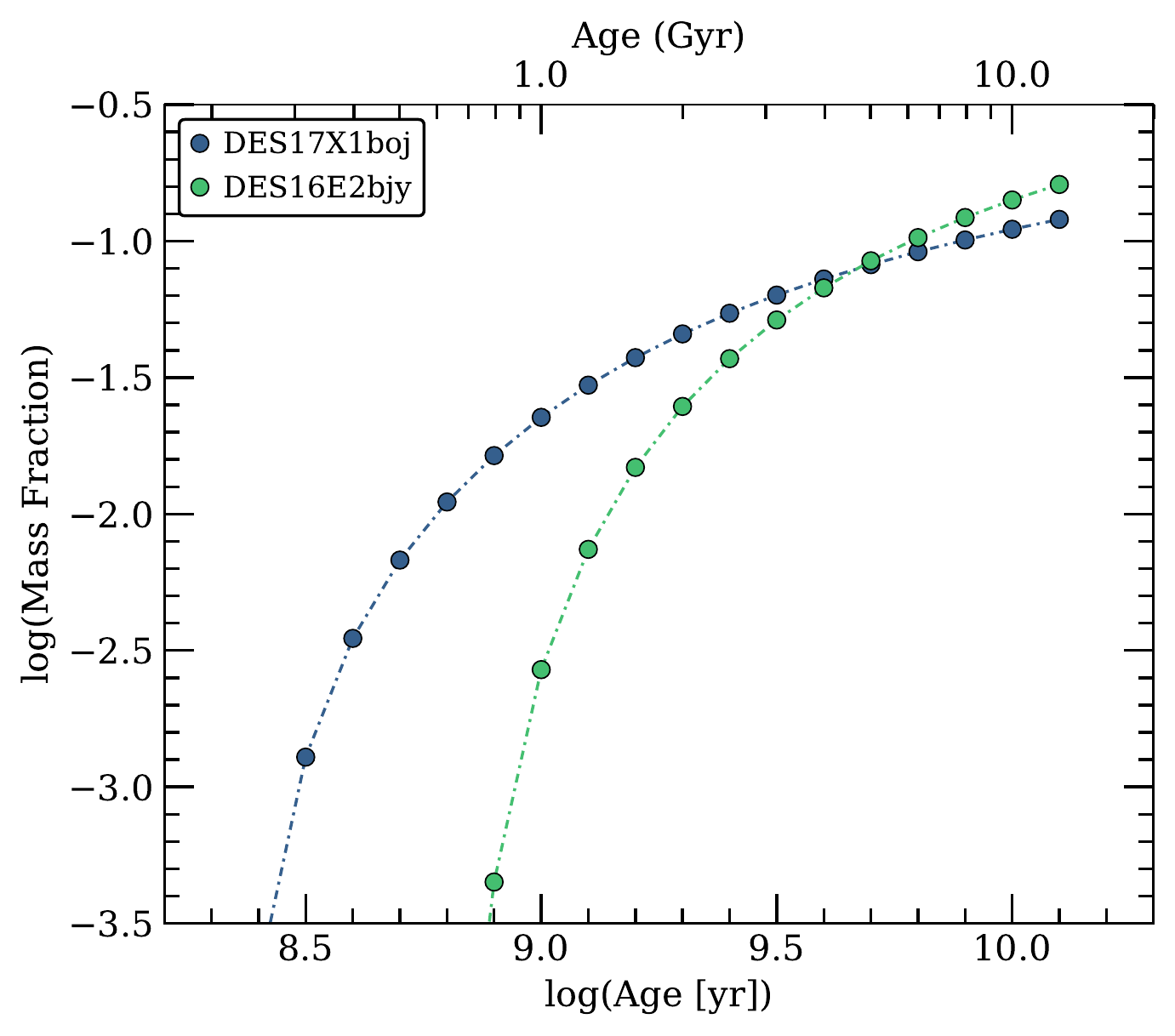}
    \caption{Star-formation history of the best-fitting models for the two transient hosts. The host of DES16E2bjy is clearly older with an average age of the stellar population of 7.6 Gyr and no significant population younger than 1 Gyr. The average age in DES17X1boj is 5.4 Gyr.}
    \label{fig:sfhs}
\end{figure}

To estimate the star-formation history (SFH) of the hosts, we fit the spectra using the pPXF spectral fitting code \citep{Cappellari2004, Cappellari2012, Cappellari2017}, based on the stellar templates provided by the MILES empirical stellar library \citep{Vazdekis2010}. pPXF simultaneously fits the continuum, stellar absorption features as well as ionised gas emission lines, providing a more robust measure of the emission-line fluxes than a continuum subtraction with a simple flat continuum. The resulting SFHs are presented in Figure \ref{fig:sfhs}. While the host of DES17X1boj consists of younger stellar populations than the host of DES16E2bjy, it still has a dominant population of older, smaller stars which is the only clear similarity between the two hosts. The presence of old populations can also be clearly seen in the spectrum itself as the spectrum is reasonably red despite the strong, ionised lines that require a young population of stars (see Figure \ref{fig:host_spectra}). The age of the young population can be estimated with the  equivalent width $W(\mathrm{H}\alpha) = 19$ Å that indicates the average age of that population to be about 10 Myr or older \citep[see e.g.][]{Kuncarayakti2016, Xiao2019}.  The average age of the continuum stellar population based on the pPFX fit is $5.4$ Gyr for the host of DES17X1boj, while for the host of DES16E2bjy it is $7.6$ Gyr with no significant population younger than 1 Gyr. 

We fit the $griz$ SEDs of the hosts with the CIGALE code \citep{Boquien2019} to determine the host galaxy masses. We use the Bruzual and Charlot stellar population models \citep{Bruzual2003} with a Salpeter initial mass function \citep{Salpeter1955}, and model the star-formation history as delayed with an optional exponential burst. The host mass of DES17X1boj was found to be $\mathrm{log(M/M_\odot)}=9.76\pm 0.16$ while for DES16E2bjy it is significantly higher with $\mathrm{log(M/M_\odot)}=10.46\pm 0.06$.

For the host of DES17X1boj we also use the flux measurements from the $\mathrm{H_\alpha}$, [SII], and [NII] lines to calculate the gas-phase metallicity \citep{Dopita2016}, which results in a slightly subsolar value of $12+\mathrm{log(O/H)} = 8.51\pm0.03$. The gas-phase metallicity for the host of DES16E2bjy can not be estimated due to lack of host galaxy emission lines. We estimated the Star-Formation Rate (SFR) for the host of DES17X1boj within the 2$\arcsec$ aperture to be $2.00\pm0.03 \cdot 10^{-2}$ M$_{\odot}$ yr$^{-1}$ based on the observed H$\alpha$ flux using the formula of \citet{Kennicutt1998}. Extrapolating this over the whole galaxy results in SFR $\sim3$ M$_{\odot}$ yr$^{-1}$, corresponding to specific star-formation rate $\mathrm{log(sSFR)}=-9.3$. While the given values are likely underestimated due to the central location of the aperture, they demonstrates that the host is moderately star-forming.

Apart from DES-SN optical photometry, the host galaxy of DES16E2bjy has also been detected in the ultraviolet (UV) regime by The Galaxy Evolution Explorer  \citep[GALEX; ][]{Martin2005} in both far UV (FUV, 1350-1750 Å in observer frame) and near UV (NUV, and 1750-2750 Å). The UV magnitudes ($\mathrm{FUV} = 22.75\pm0.07$ and $\mathrm{NUV}=22.56\pm0.05$) correspond to colors $\mathrm{FUV}-r=3.99$ and $\mathrm{NUV}-r=3.79$ indicating that the host is rather bright in UV for a such passive galaxy in the optical \citep[see e.g. ][]{Petty2013}. Assuming that all the UV emission of the galaxy comes from some residual star-formation still occurring in the galaxy we find $\mathrm{SFR}=0.14$ M$_{\odot}$ yr$^{-1}$ \citep[using the formula from][]{Kennicutt1998} and corresponding specific star-formation rate of $\mathrm{log(sSFR)}=-11.3$. While these values demonstrate that some star-formation could still be happening in the host, its level is fairly low. 

\begin{figure*}
    \centering
    \includegraphics[width=1.00\textwidth]{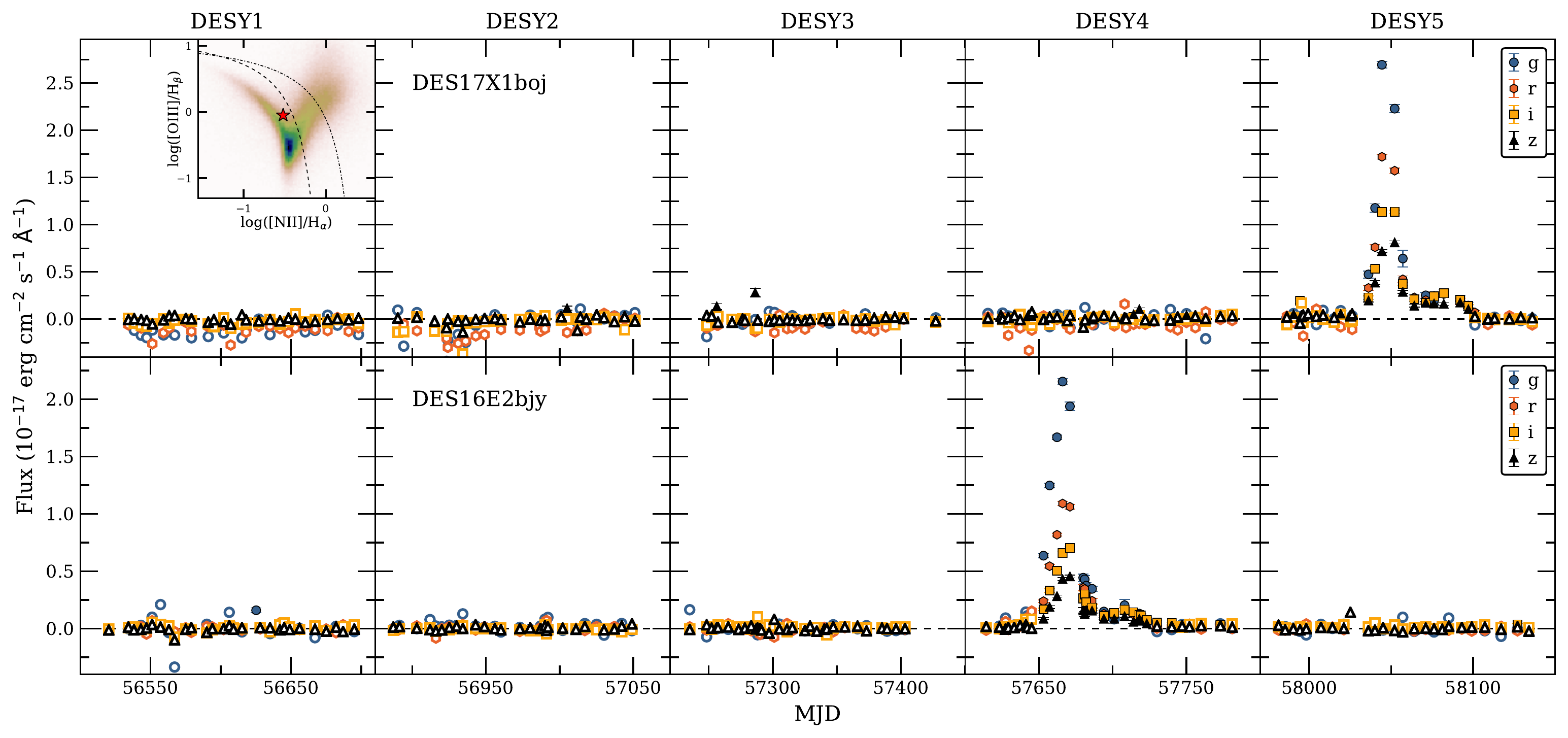}
    \caption{5-year flux light curves of DES17X1boj (top) and DES16E2bjy (bottom). Neither of the transients show significant variation outside the events themselves. A BPT diagram \citep{Baldwin1981} created with SDSS data \citep{Tremonti2004} has been plotted for DES17X1boj. The host galaxy, plotted with red star, is located below the theoretical lines from \citet{Kewley2001} and \citet{Kauffmann2003} and thus it is unlikely that the host harbours an AGN. Note that $3\sigma$ detections are plotted with filled markers and non-detections with open markers and that errors are shown only for the detections.}
    \label{fig:boj_bjy_5yr}
\end{figure*} 

As the location of DES17X1boj is consistent with the nucleus of its host galaxy it is essential to verify if the host harbours an Active Galactic Nucleus (AGN). In Figure \ref{fig:boj_bjy_5yr} (top), we show that the 5-year DES-SN light curve of the transient shows no significant variation that would be expected for an AGN. In the same Figure we also show the location of emission line flux ratios for the host galaxy on the Baldwin-Phillips-Terlevich (BPT) diagram \citep{Baldwin1981}, compared to data from SDSS \citep{Tremonti2004} and theoretical lines from \citet{Kewley2001} and \citet{Kauffmann2003}. BPT diagrams use line ratios (in this case [NII]$_\mathrm{\lambda5007}$/H$_\mathrm{\alpha}$ and [OIII]$_\mathrm{\lambda6584}$/H$_\mathrm{\beta}$) to distinguish if the line emitting region has been excited by star-formation or an AGN. Based on the diagram we see that the host of DES17X1boj is found below both of the theoretical lines, indicating that the host does not appear to have an AGN and thus the transient is unlikely to be an AGN flare. In Figure \ref{fig:boj_bjy_5yr} (bottom) we also show that the 5-year light curve of DES16E2bjy does not exhibit variation either. 

Due to the reasonably large separation of the DES16E2bjy and the centre of its assumed host galaxy ($5\arcsec$), it is possible that there is a small satellite galaxy in the vicinity of the passive galaxy that is actually the host. However, while something faint is present at the location of the transient in the DES-SN deep stacks (see Figure \ref{fig:stamps}), its measured colours match with the passive galaxy and thus it is likely associated with the putative host due to its similar stellar population. 

\section{Discussion \& Conclusions}
\label{sec:discussions}

In the sections above, we have presented our analysis of the two peculiar DES-SN transients, DES17X1boj and DES16E2bjy, and their host galaxies. Based on the relatively fast photospheric expansion velocities ($v\approx1800$ km/s for DES17X1boj and $v\approx4800$ km/s for DES16E2bjy, based on the blackbody fits) the transients are likely to be either explosive or eruptive in origin. Additionally, the blueshifted CaII absorption feature identified in the near-peak spectrum of DES17X1boj may imply that it is a SN. As the light curves of the transients evolve very similarly, it is natural to assume that DES17X1boj and DES16E2bjy are products of a similar evolutionary channel. However, the large difference in the peak luminosities and photospheric expansion velocities are difficult to describe under any single scenario.   Thus, while in the following we discuss scenarios that could explain the peculiar observational features of both of them, we emphasise that whether these two odd transients originate in similar progenitors is still unclear. 

The most striking photometric feature of DES17X1boj and DES16E2bjy is their peculiar double-peaked light curve evolution. Several CCSNe and SLSNe have been observed to have precursory bumps often credited to a short-lived shock cooling phase early in the light curve evolution and in many of these cases the pre-bumps have actually been as bright or brighter than the ``normal'' peak of the SN \citep[see e.g. SN1993J and iPTF14gqr; ][]{Wheeler1993, De2018a}. However, these pre-bump features typically last for only $\lesssim10$ d due to rapidly cooling/expanding material \citep[see e.g. DES14X3taz;][]{Smith2016}. The main peaks of the two DES-SN events last for roughly 20 days during which the photospheric radii are clearly increasing. If we assume that these peaks are powered by shock cooling in extended material, we need to explain why they last so long. One possibility could be linked to the photospheric expansion velocities that are not particularly high ($v\approx 1800$ km/s for DES17X1boj and $v\approx4800$ km/s for DES16E2bjy) and thus it is possible that extended material could stay optically thick for such a long period of time to produce the main peaks of the light curves. The secondary peak would then occur after the extended material has dissipated and an underlying slow-rising CCSN would finally emerge. However, while the secondary peak of DES16E2bjy was reasonable for CCSNe at $M_\mathrm{r}\approx-16$,  $M_\mathrm{r}\approx-13.5$ of DES17X1boj would be very faint. The faintest known CCSN to date is the Type II SN~1999br at $M=-13.77$ \citep{Pastorello2004, Anderson2014}. Additionally, having a shock cooling pre-peak that is 2-3 mag brighter than the main SN would be peculiar. Unfortunately both of the transients were too faint at the time of the second peak for spectroscopic follow-up and thus we cannot make spectroscopic comparisons at these later epochs. 

Our data pose an additional challenge to the shock cooling scenario. While the temperature evolution of DES16E2bjy is rapidly declining from the first detection \citep[as expected for rapidly heated, expanding material, see e.g. iPTF14gqr; ][]{De2018a}, the temperature of DES17X1boj increases slightly up to the peak brightness. Such an evolution can be explained with shock cooling only if the temperature gradient of the shock heated material is steep enough so that the decreasing opacity could allow us to see deeper layers of the ejecta which are still (despite adiabatic expansion) hotter than the outer layer of the photosphere was at the beginning of the light curve. Whether such a configuration is physically feasible is unclear. 

The evolution towards bluer colors and higher temperatures during the rise of a transient is not a particularly rare phenomenon itself and it can be seen in various kinds of transients. Type Ia SNe become famously bluer in their early light curve evolution \citep[see e.g.; ][]{Burns2014} as do many stripped envelope SN \citep[SESN, e.g. SN1999ex and SN2010as;][respectively]{Stritzinger2002, Folatelli2014}. Furthermore, type IIb SN2011dh mentioned earlier shows increasing photospheric temperature for the first ten days of the light curve \citep{Ergon2014}. However, this is not a feature seen only in SNe as, for example, a recent intermediate luminosity red transient (ILRT), M51 OT2019-1, evolved bluer during its rise due to destruction of circumstellar dust in the near vicinity of the very strongly reddened transient \citep{Jencson2019}.

As the spectrum of DES17X1boj shows some broad excess emission around H$\alpha$ line (see Figure \ref{fig:boj_halpha}), it is natural to compare the DES transients with type II SNe, especially given that there are some photometric similarities as well. For instance, the peak brightnesses fit well within the distribution of type II SNe \citep[$-14\gtrsim M\gtrsim -19$, see e.g.][]{Anderson2014, Richardson2014}, the majority of type II SNe have plateaus after peak that in same cases show shallow rising \citep[see e.g. SN2009N; ][]{Takats2014}, and both photometry and spectroscopy are well described with a blackbody in the beginning of the light curve. Under this scenario one would have to assume that the secondary peaks of the DES events would then be plateaus seen in SNe II. However, this is problematic due to the following reasons. The light curves drop 2 mags in 15-20 days in $r$ band (see Figure \ref{fig:lcs}) before the start of the secondary peak. Such a decline rate is significantly faster than detected for type II SNe \citep[0.9 - 8.2 mag/100 d in $V$ band;][]{Anderson2014, Gutierrez2017a}. The ``plateaus'', lasting for $\approx 20$ d, would also be on the short side of what would be expected of type IIs \citep[25 - 72 d in $V$ band;][]{Anderson2014}. Additionally, the duration of SN II plateaus appear to be correlated with peak brightness so that a shorter plateau is associated with a brighter peak magnitude \citep{Anderson2014, Galbany2016}. Thus a short plateau would be unusual for faint a transient such as DES17X1boj. Even if the spectrum of DES17X1boj shows some broad excess around H$\alpha$, some of the discussed photometric properties of the DES-SN transients do not cohere with observed type II SNe. Therefore, if they actually were type II SNe they would have to be peculiar, and given their apparent similarity to each other they might be first examples of a strange kind of type II SNe. 

Another interesting type of transient to compare DES17X1boj and DES16E2bjy with are the SN impostors. As shown in Figure \ref{fig:lcs_comp}, SN2009ip has a short phase of re-brightening around the same phase as the secondary peak of the DES-SN transients, and its peak brightness ($M_\mathrm{V}=-17.7$; see e.g. \citealt{Fraser2013}) is similar to DES16E2bjy. However, several other features distinguish our double-peaked DES-SN transients from the SN impostors. While SN2009ip does show rebrightening, its light curve evolution is clearly different from the DES transients. Additionally, other impostor candidates such as SN2015bh \citep[see e.g.][]{Elias-Rosa2016} and SN2016bdu \citep{Pastorello2018} have very similar light curves with SN2009ip, but do not exhibit rebrightening. Furthermore, our photometric data also constrains the long term variability of DES17X1boj to a level below what was seen in SN2009ip \citep{Pastorello2013} and SN2016bdu \citep{Pastorello2018} in the years before the brightest event ($M_\mathrm{V}$ in range -13 to -14). For the more distant event DES16E2bjy, such outbursts would have been below our detection threshold.  Regarding to spectroscopic data, the impostors exhibit strong, narrow hydrogen and helium lines around peak brightness \citep[see e.g. ][]{Fraser2013, Mauerhan2013, Pastorello2013}. No such features are seen in either of the DES transients (see Figures \ref{fig:spectra_boj} and \ref{fig:spectra_bjy}), but it is possible that the lines are hidden in the noise.To investigate this we estimated the limiting equivalent width (EW) for a Gaussian-shaped narrow H$\alpha$ line with $v_\mathrm{FWHM}=500$ km/s in our spectra. Hydrogen lines with similar widths are often seen in both SN impostors \citep{Smith2011} and in type IIn SNe \citep{Taddia2013} where $v_\mathrm{FWHM}\sim 100 - 1000$ km/s are typically measured. For the given configuration we found limits of EW $\lesssim5$ Å for DES17X1boj and EW $\lesssim14$ Å for DES16E2bjy. In the case of of type IIn SNe the line strengths are typically measured in several tens to hundreds of Ångstroms \citep[EW$\gtrsim 40$ Å; ][]{Smith2014}, and thus it is unlikely that narrow H$\alpha$ lines are hiding in the spectra. Due to both photometric and spectroscopic differences it is unlikely that DES17X1boj and DES16E2bjy are events similar to SN impostors.


Luminous Red Novae (LRNe) exhibit long-term light curve evolution often with several peaks like is the case for SNhunt248 \citep[e.g. ][]{Kankare2015} shown in Figure \ref{fig:lcs_comp}. LRNe are also photometrically very inhomogeneous class of transients and reach luminosities of $M_\mathrm{V}\gtrsim-15$ \citep[e.g. ][]{Pastorello2019} and therefore at least DES17X1boj could be a member of this class. However, the light curve evolution of LRN is typically significantly slower than what is seen in the DES transients and DES16E2bjy is approximately three magnitudes brighter than any classified LRNe.In regards to spectra, many LRNe exhibit strong, narrow lines in their spectra throughout their light curves \citep[see e.g.][]{Pastorello2019}, but not all of them. For instance, for V838 Mon narrow H$_\mathrm{\alpha}$ emission was present 5 days after peak brightness \citep{Smith2011}, but the line was not visible in the decline phase 30 to 60 days post-peak \citep{Smith2016a}. Thus, while the absence of narrow emission lines in the spectra of the DES transients (see Figures \ref{fig:spectra_boj} and \ref{fig:spectra_bjy}) cannot be used to rule out the LRNe as a possible scenario, different light curve evolution timescales and high peak brightness of DES16E2bjy makes it unlikely.

One topic that could give us insight on the origin of the DES-SN transients is where they occurred in their host galaxies. The event location is one of the most constraining observational features about DES16E2bjy: it is found in the outskirts of a passive galaxy. If this galaxy is truly the host of DES16E2bjy it would  disfavour scenarios related to progenitors with massive stars as only few CCSNe have been associated with passive galaxies (see e.g. Type II SNe Abell399\_11\_19\_0 and SN2016hil and Ibn SN PS1-12sk ; \citealt{Graham2012}, \citealt{Irani2019} and \citealt{Hosseinzadeh2019}, respectively). This is almost contradictory to the fact that the photometric and spectroscopic data appears to be well described with a blackbody, which typically requires a significant amount of material that can be shock heated and thus it is a feature often seen in CCSNe. Therefore, we conclude one of the following must be true: either there has been a small amount of recent star-formation in the host galaxy leading to this peculiar transient, or this is a new type of transient that originates in an environment with old stellar populations. As discussed in Section \ref{sec:hostgal}, the $\mathrm{UV}-r$ colors are higher than expected from from early type galaxies, potentially suggesting that residual star-formation is still occurring in the galaxy \citep[see e.g.][]{Petty2013}.  Therefore, it is not possible to conclusively determine if DES16E2bjy is a product of an old stellar population. DES17X1boj, on the other hand, is associated with a star-forming galaxy with a non-negligible old stellar population, so the transient could originate from either young or old stellar populations. 

As DES16E2bjy appears to be associated with a passive galaxy, one also has to consider a thermonuclear origin for it. However, as seen in Figure \ref{fig:lcs_comp}, its light curve shape is completely different when compared with prototypical type Ia SN2011fe, even if its absolute magnitude is loosely within the Ia distribution \citep[$-18\gtrsim M\gtrsim -20$, see e.g.][]{Richardson2014}. Comparison with type Iax SNe, the largest class of peculiar thermonuclear SNe also known as SN2002cx-like \citep{Li2003, Jha2006, Foley2013}, yields a similar result. While their peak magnitudes are lower than for ``normal’’ type Ia SNe \citep[$-13\gtrsim M\gtrsim -19$, see e.g.][]{Jha2017}, they have no stronger secondary peaks powered by FeIII recombination seen in type Ia SNe and therefore they are clearly different than DES16E2bjy. Additionally, the color of SNe Iax is significantly redder than for DES16E2bjy in the beginning of the light curve \citep[$B-V \gtrsim 0.0; $][]{Foley2013}. Due to the different light curve shapes and inconsistent colors, it is unlikely that thermonuclear SN alone can be responsible for DES16E2bjy and even less for DES17X1boj due to its lower peak magnitude. 

It is possible that the DES-SN transients are thermonuclear with addition of some other power source, for instance interaction with circumstellar material (CSM). Under this scenario the thermonuclear explosion cannot be a standard type Ia SN as they are too bright, so it would have to be a peculiar one such as type Iax. However, as these appear to be too red in the beginning of the light curve, the interaction would have to be such that it causes the beginning of the light curve to be bluer and still produces the secondary peak. While no narrow emission lines are present in the peak spectra of the DES transients, the EW of narrow H$\alpha$ emission in the interacting type Ia SNe (Ia-CSM) is not statistically different than for IIn SNe \citep{Silverman2013}. Thus based on the limits of the emission we estimated, it is unlikely that there is a significant amount of interaction at the time of peak brightness. The presence of interaction cannot be ruled out during the secondary peaks, as there was no spectral coverage during these epochs. However, due to the needed strong effect of interaction on the light curves of the DES transients, this particular scenario of thermonuclear SNe with CSM interaction sounds contrived. Therefore, we consider it unlikely but it is possible that some other power source in addition to peculiar thermonuclear SNe could produce these transients. 

DES17X1boj occurred in a star-forming galaxy so it is possible that at least some of the brightness difference between the two transients could be explained by host galaxy extinction. However, the blue bands (especially $g$) are bright and persistent throughout the light curves (see Figure \ref{fig:lcs}) and the the blackbody fits are very good in the beginning of the light curve (see Figure \ref{fig:bb_boj}), neither of which would be expected in case of strong extinction. Thus, while it is likely that there is some extinction in the host galaxy, meaning DES17X1boj is intrinsically bluer and brighter, it does not seem probable that the difference in brightness between the two transients is caused by that. 

As both DES-SN events were found at relatively low redshifts and as DES16E2bjy is reasonably bright at peak, it is interesting that no transients with similar light curve evolution have been identified in the literature, especially if we want to assume that they are powered by same physical mechanism. However, as the main characterising feature is the faint, secondary peak seen in all four $griz$ bands some of these transients may have gone by unrecognised. This argument can be put in context with the example of Ca-rich transients described by absolute magnitudes of -15 to -16.5. Even though the first such event was published nearly a decade ago \citep[SN2005E;][]{Perets2010} their total number has stayed low ($\sim$10) even if their volumetric rate in local universe has been estimated to be significant fraction of the Ia rate \citep[$\gtrsim10$\%; ][]{Perets2010, Frohmaier2018}. This demonstrates that even if transients such as DES17X1boj or DES16E2bjy were relatively common, the faintness of the secondary peak makes them difficult to separate from the general SN population. 

Here we have presented our analysis of the two peculiar double-peaked DES-SN transients, DES17X1boj and DES16E2bjy. As discussed earlier in this section, any standard SN scenario creates more conflicts than it solves. While early photometric and spectroscopic data would lean towards an origin related to CCSNe, it is difficult to explain why such a peculiar CCSN would be found in a passive galaxy. While this conflict might convince one that these transients must then be thermonuclear, their secondary peaks cannot be produced via means of FeIII recombination like is the case for thermonuclear SNe, making this scenario contrived as well. It is also entirely possible that, despite nearly identical light curves and the fact that no similar transients are found in the literature, the events are not physically similar to each other and are in fact produced via different progenitor channels. More similar transients need to be discovered to verify if they are truly similar, and if so, what the progenitors are like.

\section*{Acknowledgements}
We thank the anonymous referee for the helpful feedback and Andrea Pastorello for fruitful discussion. We acknowledge support from EU/FP7 ERC grant 615929, and STFC grant ST/R000506/1. L.G. was funded by the European Union's Horizon 2020 research and innovation programme under the Marie Sk\l{}odowska-Curie grant agreement No. 839090.

Some of the observations reported in this paper were obtained with the Southern African Large Telescope (SALT) and the Galaxy and Mass Assembly programme (GAMA). GAMA is a joint European-Australasian project based around a spectroscopic campaign using the Anglo-Australian Telescope. The GAMA input catalogue is based on data taken from the Sloan Digital Sky Survey and the UKIRT Infrared Deep Sky Survey. Complementary imaging of the GAMA regions is being obtained by a number of independent survey programmes including GALEX MIS, VST KiDS, VISTA VIKING, WISE, Herschel-ATLAS, GMRT and ASKAP providing UV to radio coverage. GAMA is funded by the STFC (UK), the ARC (Australia), the AAO, and the participating institutions. The GAMA website is http://www.gama-survey.org/. 

Funding for the DES Projects has been provided by the U.S. Department of Energy, the U.S. National Science Foundation, the Ministry of Science and Education of Spain, 
the Science and Technology Facilities Council of the United Kingdom, the Higher Education Funding Council for England, the National Center for Supercomputing 
Applications at the University of Illinois at Urbana-Champaign, the Kavli Institute of Cosmological Physics at the University of Chicago, 
the Center for Cosmology and Astro-Particle Physics at the Ohio State University,
the Mitchell Institute for Fundamental Physics and Astronomy at Texas A\&M University, Financiadora de Estudos e Projetos, 
Funda{\c c}{\~a}o Carlos Chagas Filho de Amparo {\`a} Pesquisa do Estado do Rio de Janeiro, Conselho Nacional de Desenvolvimento Cient{\'i}fico e Tecnol{\'o}gico and 
the Minist{\'e}rio da Ci{\^e}ncia, Tecnologia e Inova{\c c}{\~a}o, the Deutsche Forschungsgemeinschaft and the Collaborating Institutions in the Dark Energy Survey. 

The Collaborating Institutions are Argonne National Laboratory, the University of California at Santa Cruz, the University of Cambridge, Centro de Investigaciones Energ{\'e}ticas, 
Medioambientales y Tecnol{\'o}gicas-Madrid, the University of Chicago, University College London, the DES-Brazil Consortium, the University of Edinburgh, 
the Eidgen{\"o}ssische Technische Hochschule (ETH) Z{\"u}rich, 
Fermi National Accelerator Laboratory, the University of Illinois at Urbana-Champaign, the Institut de Ci{\`e}ncies de l'Espai (IEEC/CSIC), 
the Institut de F{\'i}sica d'Altes Energies, Lawrence Berkeley National Laboratory, the Ludwig-Maximilians Universit{\"a}t M{\"u}nchen and the associated Excellence Cluster Universe, 
the University of Michigan, the National Optical Astronomy Observatory, the University of Nottingham, The Ohio State University, the University of Pennsylvania, the University of Portsmouth, 
SLAC National Accelerator Laboratory, Stanford University, the University of Sussex, Texas A\&M University, and the OzDES Membership Consortium.

Based in part on observations at Cerro Tololo Inter-American Observatory, National Optical Astronomy Observatory, which is operated by the Association of 
Universities for Research in Astronomy (AURA) under a cooperative agreement with the National Science Foundation. Based in part on data acquired at the Anglo-Australian Telescope, under program A/2013B/012. We acknowledge the traditional owners of the land on which the AAT stands, the Gamilaraay people, and pay our respects to elders past and present.

The DES data management system is supported by the National Science Foundation under Grant Numbers AST-1138766 and AST-1536171.
The DES participants from Spanish institutions are partially supported by MINECO under grants AYA2015-71825, ESP2015-66861, FPA2015-68048, SEV-2016-0588, SEV-2016-0597, and MDM-2015-0509, 
some of which include ERDF funds from the European Union. IFAE is partially funded by the CERCA program of the Generalitat de Catalunya.
Research leading to these results has received funding from the European Research
Council under the European Union's Seventh Framework Program (FP7/2007-2013) including ERC grant agreements 240672, 291329, and 306478.
We  acknowledge support from the Brazilian Instituto Nacional de Ci\^encia
e Tecnologia (INCT) e-Universe (CNPq grant 465376/2014-2).

This manuscript has been authored by Fermi Research Alliance, LLC under Contract No. DE-AC02-07CH11359 with the U.S. Department of Energy, Office of Science, Office of High Energy Physics. The United States Government retains and the publisher, by accepting the article for publication, acknowledges that the United States Government retains a non-exclusive, paid-up, irrevocable, world-wide license to publish or reproduce the published form of this manuscript, or allow others to do so, for United States Government purposes.

This research used resources of the National Energy Research Scientific Computing Center (NERSC), a U.S. Department of Energy Office of Science User Facility operated under Contract No. DE-AC02-05CH11231.




\bibliographystyle{mnras}
\bibliography{bib} 




\appendix
\section{Blackbody fits}

\begin{figure*}
    \centering
    \includegraphics[width=0.98\textwidth]{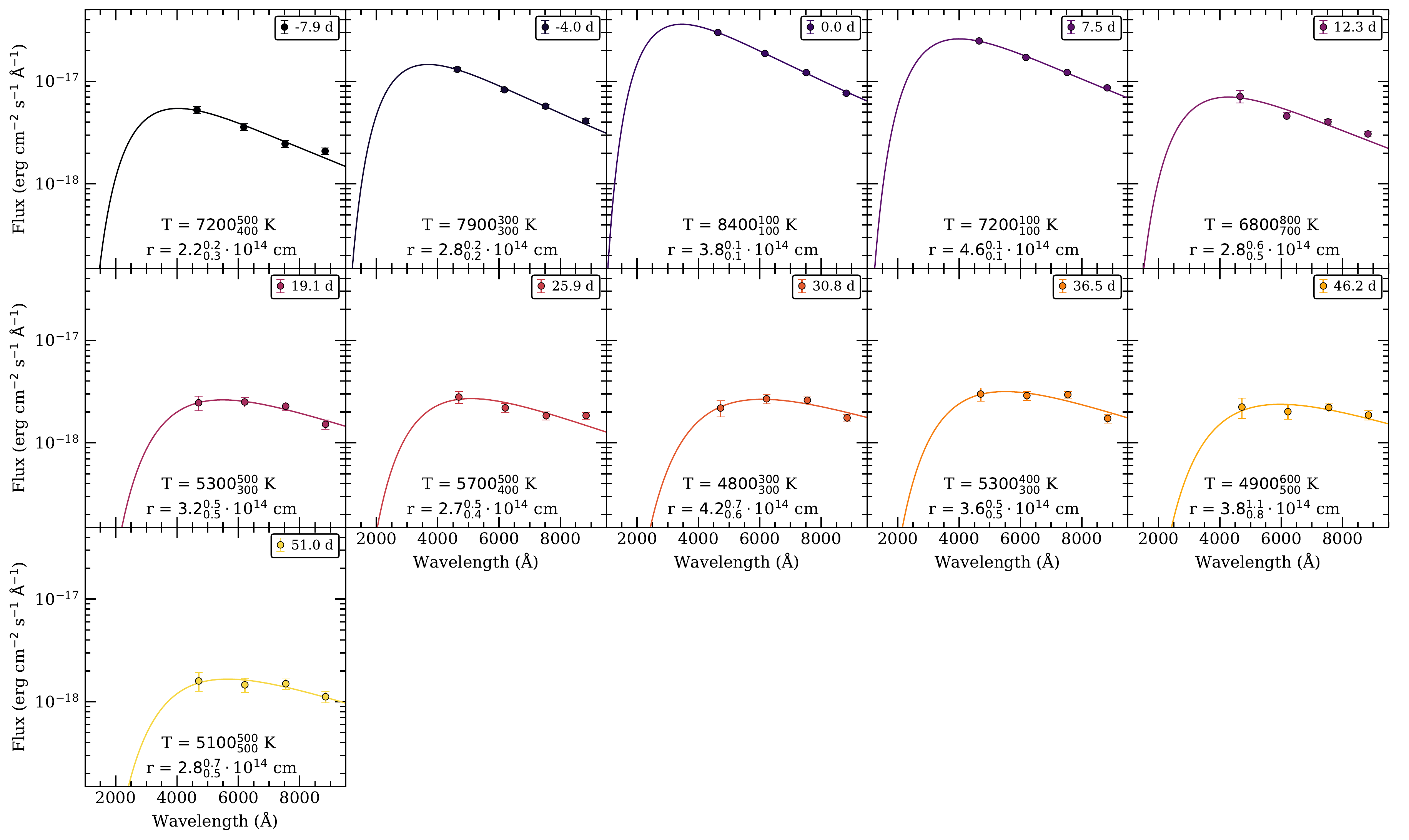}
    \caption{Blackbody fits to photometry of DES17X1boj for every epoch with $griz$ detections. Best-fitting temperatures and radii are given with  1$\sigma$ uncertainties.}
    \label{fig:bb_boj}
\end{figure*}

\begin{figure*}
    \centering
    \includegraphics[width=0.98\textwidth]{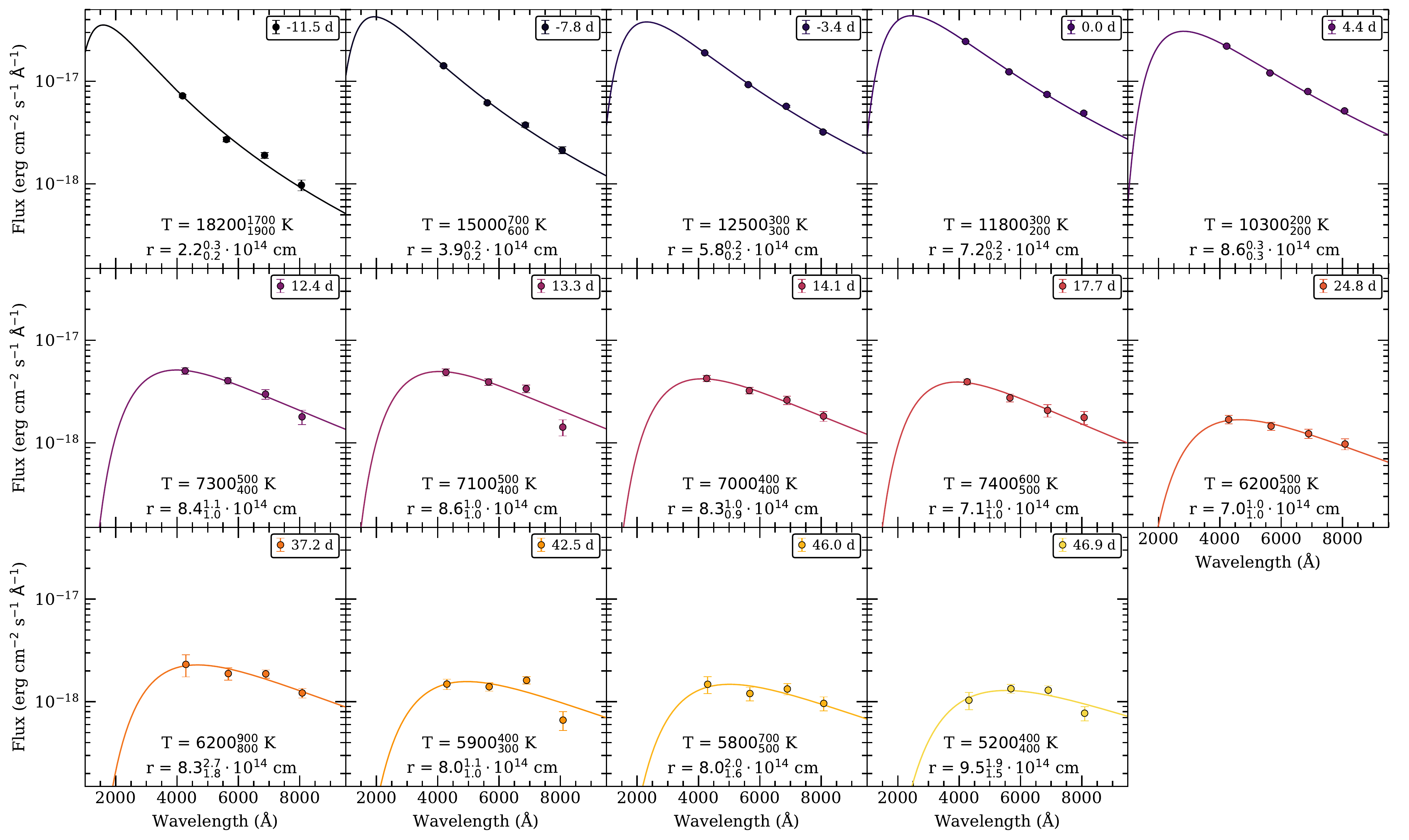}
    \caption{Blackbody fits to photometry of DES16E2bjy for every epoch with $griz$ detections. Best-fitting temperatures and radii are given with  1$\sigma$ uncertainties.}
    \label{fig:bb_bjy}
\end{figure*}

The blackbody fits to the data of DES17X1boj and DES16E2bjy are shown in Figures \ref{fig:bb_boj} and \ref{fig:bb_bjy}, respectively.

\label{app:bb_fits}

\clearpage
\input DES-2019-0495_affiliations_only_final2.tex


\bsp	
\label{lastpage}
\end{document}

%% file: DES-2019-0495_author_list_only_final2.tex
\author[DES Collaboration]{
\parbox{\textwidth}{
\Large
M.~Pursiainen,$^{1}$
C.~P.~Guti\'errez,$^{1}$
P.~Wiseman,$^{1}$
M.~Childress,$^{1}$
M.~Smith,$^{1}$
C.~Frohmaier,$^{2}$
C.~Angus,$^{1}$
N.~Castro Segura,$^{1}$
L.~Kelsey,$^{1}$
M.~Sullivan,$^{1}$
L.~Galbany,$^{3}$
P.~Nugent,$^{4}$
B.~A.~Bassett,$^{5,6}$
D.~Brout,$^{7}$
D.~Carollo,$^{8}$
C.~B.~D'Andrea,$^{9}$
T.~M.~Davis,$^{10}$
R.~J.~Foley,$^{11}$
M.~Grayling,$^{1}$
S.~R.~Hinton,$^{10}$
C.~Inserra,$^{12}$
R.~Kessler,$^{13,14}$
G.~F.~Lewis,$^{15}$
C.~Lidman,$^{16},$
E.~Macaulay,$^{2}$
M.~March,$^{7}$
A.~M\"oller,$^{17}$
T.~M\"uller,$^{1}$
D.~Scolnic,$^{18},$
N.~E.~Sommer,$^{16}$
E.~Swann,$^{2}$
B.~P.~Thomas,$^{19}$
B.~E.~Tucker,$^{16}$
M.~Vincenzi,$^{2}$
T.~M.~C.~Abbott,$^{20}$
S.~Allam,$^{21}$
J.~Annis,$^{21}$
S.~Avila,$^{22}$
E.~Bertin,$^{23,24}$
D.~Brooks,$^{25}$
E.~Buckley-Geer,$^{21}$
D.~L.~Burke,$^{26,27}$
A.~Carnero~Rosell,$^{28,29}$
M.~Carrasco~Kind,$^{30,31}$
L.~N.~da Costa,$^{29,32}$
J.~De~Vicente,$^{28}$
S.~Desai,$^{33}$
H.~T.~Diehl,$^{21}$
P.~Doel,$^{25}$
T.~F.~Eifler,$^{34,35}$
S.~Everett,$^{11}$
B.~Flaugher,$^{21}$
J.~Frieman,$^{21,14}$
J.~Garc\'ia-Bellido,$^{22}$
E.~Gaztanaga,$^{36,37}$
D.~W.~Gerdes,$^{38,39}$
D.~Gruen,$^{40,26,27}$
R.~A.~Gruendl,$^{30,31}$
J.~Gschwend,$^{29,32}$
G.~Gutierrez,$^{21}$
D.~L.~Hollowood,$^{11}$
K.~Honscheid,$^{41,42}$
D.~J.~James,$^{43}$
A.~G.~Kim,$^{4}$
E.~Krause,$^{34}$
K.~Kuehn,$^{44,45}$
M.~A.~G.~Maia,$^{29,32}$
J.~L.~Marshall,$^{46}$
F.~Menanteau,$^{30,31}$
R.~Miquel,$^{47,48}$
R.~L.~C.~Ogando,$^{29,32}$
A.~Palmese,$^{21,14}$
F.~Paz-Chinch\'{o}n,$^{30,31}$
A.~A.~Plazas,$^{49}$
A.~Roodman,$^{26,27}$
E.~Sanchez,$^{28}$
V.~Scarpine,$^{21}$
M.~Schubnell,$^{39}$
S.~Serrano,$^{36,37}$
I.~Sevilla-Noarbe,$^{28}$
E.~Suchyta,$^{50}$
M.~E.~C.~Swanson,$^{31}$
G.~Tarle,$^{39}$
W.~Wester$^{21}$
\begin{center} (DES Collaboration) \end{center}
}
\vspace{0.4cm}
\\
\parbox{\textwidth}{Affiliations are listed at the end of the paper}
}

%% file: DES-2019-0495_affiliations_only_final2.tex
\parbox{\textwidth}{
$^{1}$ School of Physics and Astronomy, University of Southampton,  Southampton, SO17 1BJ, UK\\
$^{2}$ Institute of Cosmology and Gravitation, University of Portsmouth, Portsmouth, PO1 3FX, UK\\
$^{3}$ Departamento de F\'isica Te\'orica y del Cosmos, Universidad de Granada, E-18071 Granada, Spain\\
$^{4}$ Lawrence Berkeley National Laboratory, 1 Cyclotron Road, Berkeley, CA 94720, USA\\
$^{5}$ African Institute for Mathematical Sciences, 6 Melrose Road, Muizenberg, 7945, South Africa\\
$^{6}$ South African Astronomical Observatory, P.O.Box 9, Observatory 7935, South Africa\\
$^{7}$ Department of Physics and Astronomy, University of Pennsylvania, Philadelphia, PA 19104, USA\\
$^{8}$ INAF, Astrophysical Observatory of Turin, I-10025 Pino Torinese, Italy\\
$^{9}$ Department of Physics and Astronomy, University of Pennsylvania, Philadelphia, PA 19104, USA\\
$^{10}$ School of Mathematics and Physics, University of Queensland,  Brisbane, QLD 4072, Australia\\
$^{11}$ Santa Cruz Institute for Particle Physics, Santa Cruz, CA 95064, USA\\
$^{12}$ School of Physics \& Astronomy, Cardiff University, Queens Buildings, The Parade, Cardiff, CF24 3AA, UK\\
$^{13}$ Department of Astronomy and Astrophysics, University of Chicago, Chicago, IL 60637, USA\\
$^{14}$ Kavli Institute for Cosmological Physics, University of Chicago, Chicago, IL 60637, USA\\
$^{15}$ Sydney Institute for Astronomy, School of Physics, A28, The University of Sydney, NSW 2006, Australia\\
$^{16}$ The Research School of Astronomy and Astrophysics, Australian National University, ACT 2601, Australia\\
$^{17}$ Universit\'e Clermont Auvergne, CNRS/IN2P3, LPC, F-63000 Clermont-Ferrand, France\\
$^{18}$ Department of Physics, Duke University Durham, NC 27708, USA\\
$^{19}$ Department of Astronomy, University of Texas at Austin, Austin, TX,USA\\
$^{20}$ Cerro Tololo Inter-American Observatory, National Optical Astronomy Observatory, Casilla 603, La Serena, Chile\\
$^{21}$ Fermi National Accelerator Laboratory, P. O. Box 500, Batavia, IL 60510, USA\\
$^{22}$ Instituto de Fisica Teorica UAM/CSIC, Universidad Autonoma de Madrid, 28049 Madrid, Spain\\
$^{23}$ CNRS, UMR 7095, Institut d'Astrophysique de Paris, F-75014, Paris, France\\
$^{24}$ Sorbonne Universit\'es, UPMC Univ Paris 06, UMR 7095, Institut d'Astrophysique de Paris, F-75014, Paris, France\\
$^{25}$ Department of Physics \& Astronomy, University College London, Gower Street, London, WC1E 6BT, UK\\
$^{26}$ Kavli Institute for Particle Astrophysics \& Cosmology, P. O. Box 2450, Stanford University, Stanford, CA 94305, USA\\
$^{27}$ SLAC National Accelerator Laboratory, Menlo Park, CA 94025, USA\\
$^{28}$ Centro de Investigaciones Energ\'eticas, Medioambientales y Tecnol\'ogicas (CIEMAT), Madrid, Spain\\
$^{29}$ Laborat\'orio Interinstitucional de e-Astronomia - LIneA, Rua Gal. Jos\'e Cristino 77, Rio de Janeiro, RJ - 20921-400, Brazil\\
$^{30}$ Department of Astronomy, University of Illinois at Urbana-Champaign, 1002 W. Green Street, Urbana, IL 61801, USA\\
$^{31}$ National Center for Supercomputing Applications, 1205 West Clark St., Urbana, IL 61801, USA\\
$^{32}$ Observat\'orio Nacional, Rua Gal. Jos\'e Cristino 77, Rio de Janeiro, RJ - 20921-400, Brazil\\
$^{33}$ Department of Physics, IIT Hyderabad, Kandi, Telangana 502285, India\\
$^{34}$ Department of Astronomy/Steward Observatory, University of Arizona, 933 North Cherry Avenue, Tucson, AZ 85721-0065, USA\\
$^{35}$ Jet Propulsion Laboratory, California Institute of Technology, 4800 Oak Grove Dr., Pasadena, CA 91109, USA\\
$^{36}$ Institut d'Estudis Espacials de Catalunya (IEEC), 08034 Barcelona, Spain\\
$^{37}$ Institute of Space Sciences (ICE, CSIC),  Campus UAB, Carrer de Can Magrans, s/n,  08193 Barcelona, Spain\\
$^{38}$ Department of Astronomy, University of Michigan, Ann Arbor, MI 48109, USA\\
$^{39}$ Department of Physics, University of Michigan, Ann Arbor, MI 48109, USA\\
$^{40}$ Department of Physics, Stanford University, 382 Via Pueblo Mall, Stanford, CA 94305, USA\\
$^{41}$ Center for Cosmology and Astro-Particle Physics, The Ohio State University, Columbus, OH 43210, USA\\
$^{42}$ Department of Physics, The Ohio State University, Columbus, OH 43210, USA\\
$^{43}$ Center for Astrophysics $\vert$ Harvard \& Smithsonian, 60 Garden Street, Cambridge, MA 02138, USA\\
$^{44}$ Australian Astronomical Optics, Macquarie University, North Ryde, NSW 2113, Australia\\
$^{45}$ Lowell Observatory, 1400 Mars Hill Rd, Flagstaff, AZ 86001, USA\\
$^{46}$ George P. and Cynthia Woods Mitchell Institute for Fundamental Physics and Astronomy, and Department of Physics and Astronomy, Texas A\&M University, College Station, TX 77843,  USA\\
$^{47}$ Instituci\'o Catalana de Recerca i Estudis Avan\c{c}ats, E-08010 Barcelona, Spain\\
$^{48}$ Institut de F\'{\i}sica d'Altes Energies (IFAE), The Barcelona Institute of Science and Technology, Campus UAB, 08193 Bellaterra (Barcelona) Spain\\
$^{49}$ Department of Astrophysical Sciences, Princeton University, Peyton Hall, Princeton, NJ 08544, USA\\
$^{50}$ Computer Science and Mathematics Division, Oak Ridge National Laboratory, Oak Ridge, TN 37831\\
}